\documentclass[a4paper,11pt]{article}

\usepackage[colorlinks=true]{hyperref}
\usepackage[margin=1.5cm]{geometry}
\usepackage{authblk}
\usepackage{graphicx}
\usepackage{threeparttable}
\usepackage{booktabs}
\usepackage{amsmath}
\usepackage{amssymb}
\usepackage{wasysym}
\begin{document}
\title{PandaX-xT -- a Deep Underground Multi-ten-tonne Liquid Xenon Observatory}

\date{October 31, 2024}
\author[2]{Abdusalam Abdukerim}
\author[2]{Zihao Bo}
\author[2]{Wei Chen}
\author[1,3]{Xun Chen}
\author[4]{Chen Cheng}
\author[5]{Zhaokan Cheng}
\author[1]{Xiangyi Cui}
\author[6]{Yingjie Fan}
\author[7]{Deqing Fang}
\author[8,9,10,11]{Lisheng Geng}
\author[2]{Karl Giboni}
\author[2]{Linhui Gu}
\author[8]{Xunan Guo}
\author[12]{Xuyuan Guo}
\author[8]{Zhichao Guo}
\author[1]{Chencheng Han}
\author[2]{Ke Han}
\author[2]{Changda He}
\author[10]{Jinrong He}
\author[2]{Di Huang}
\author[2]{Junting Huang}
\author[2]{Zhou Huang}
\author[3]{Ruquan Hou}
\author[13]{Yu Hou}
\author[14]{Xiangdong Ji}
\author[13]{Yonglin Ju}
\author[2]{Chenxiang Li}
\author[4]{Jiafu Li}
\author[12]{Mingchuan Li}
\author[12,1]{Shuaijie Li}
\author[5]{Tao Li}
\author[15,16]{Qing Lin}
\author[1,2,3]{Jianglai Liu\thanks{spokesperson: jianglai.liu@sjtu.edu.cn}}
\author[13]{Congcong Lu}
\author[17,18]{Xiaoying Lu}
\author[19]{Lingyin Luo}
\author[16]{Yunyang Luo}
\author[2]{Wenbo Ma}
\author[7]{Yugang Ma}
\author[19]{Yajun Mao}
\author[2,3]{Yue Meng}
\author[2]{Xuyang Ning}
\author[17,18]{Binyu Pang}
\author[12]{Ningchun Qi}
\author[2]{Zhicheng Qian}
\author[17,18]{Xiangxiang Ren}
\author[17,18]{Nasir Shaheed}
\author[2]{Xiaofeng Shang}
\author[20]{Xiyuan Shao}
\author[8]{Guofang Shen}
\author[2]{Lin Si}
\author[12]{Wenliang Sun}
\author[2,3]{Yi Tao}
\author[17,18]{Anqing Wang}
\author[17,18]{Meng Wang}
\author[7]{Qiuhong Wang}
\author[2,21]{Shaobo Wang}
\author[19]{Siguang Wang}
\author[5,4]{Wei Wang}
\author[13]{Xiuli Wang}
\author[2]{Xu Wang}
\author[2,3,1]{Zhou Wang}
\author[5]{Yuehuan Wei}
\author[4]{Mengmeng Wu}
\author[2]{Weihao Wu}
\author[2]{Yuan Wu}
\author[2]{Mengjiao Xiao}
\author[4]{Xiang Xiao}
\author[1]{Binbin Yan}
\author[22]{Xiyu Yan}
\author[2]{Yong Yang}
\author[20]{Chunxu Yu}
\author[2]{Ying Yuan}
\author[7]{Zhe Yuan}
\author[2]{Youhui Yun}
\author[2]{Xinning Zeng}
\author[2]{Minzhen Zhang}
\author[12]{Peng Zhang}
\author[2]{Shibo Zhang}
\author[4]{Shu Zhang}
\author[2,3]{Tao Zhang}
\author[1]{Wei Zhang}
\author[17,18]{Yang Zhang}
\author[17,18]{Yingxin Zhang}
\author[1]{Yuanyuan Zhang}
\author[2,3]{Li Zhao}
\author[12]{Jifang Zhou}
\author[2,3]{Ning Zhou}
\author[8]{Xiaopeng Zhou}
\author[10]{Yong Zhou}
\author[2]{Yubo Zhou}
\author[2]{Zhizhen Zhou}

\affil[ ]{(PandaX Collaboration)}
\affil[1]{New Cornerstone Science Laboratory, Tsung-Dao Lee Institute, Shanghai Jiao Tong University, Shanghai, 200240, China}
\affil[2]{School of Physics and Astronomy, Shanghai Jiao Tong University, Key Laboratory for Particle Astrophysics 
and Cosmology (MoE), 
 Shanghai Key Laboratory for Particle Physics and Cosmology, Shanghai 200240, China}
\affil[3]{Shanghai Jiao Tong University Sichuan Research Institute, Chengdu 610213, China}
\affil[4]{School of Physics, Sun Yat-Sen University, Guangzhou 510275, China}
\affil[5]{Sino-French Institute of Nuclear Engineering and Technology, Sun Yat-Sen University, Zhuhai, 519082, China}
\affil[6]{Department of Physics, Yantai University, Yantai 264005, China}
\affil[7]{Key Laboratory of Nuclear Physics and Ion-beam Application (MOE), Institute of Modern Physics, Fudan University,Shanghai 200433, China}
\affil[8]{School of Physics, Beihang University, Beijing 102206, China}
\affil[9]{Peng Huanwu Collaborative Center for Research and Education, Beihang University, Beijing 100191, China}
\affil[10]{Beijing Key Laboratory of Advanced Nuclear Materials and Physics, Beihang University, Beijing 102206, China}
\affil[11]{Southern Center for Nuclear-Science Theory (SCNT), Institute of Modern Physics, Chinesen Academy of Sciences,Huizhou 516000, China}
\affil[12]{Yalong River Hydropower Development Company, Ltd., Chengdu 610051, China}
\affil[13]{School of Mechanical Engineering, Shanghai Jiao Tong University, Shanghai 200240, China}
\affil[14]{Department of Physics, University of Maryland, College Park, Maryland 20742, USA}
\affil[15]{State Key Laboratory of Particle Detection and Electronics, University of Science and Technology of China, Hefei 230026, China}
\affil[16]{Department of Modern Physics, University of Science and Technology of China, Hefei 230026, China}
\affil[17]{Research Center for Particle Science and Technology, Institute of Frontier and Interdisciplinary Science, 
Shandong University, Qingdao 266237, Shandong, China}
\affil[18]{Key Laboratory of Particle Physics and Particle Irradiation of Ministry of Education, Shandong University, 
Qingdao 266237, Shandong, China}
\affil[19]{State Key Laboratory of Nuclear Physics and Technology, School of Physics, Peking University, Beijing 100871, China}
\affil[20]{School of Physics, Nankai University, Tianjin 300071, China}
\affil[21]{SJTU Paris Elite Institute of Technology, Shanghai Jiao Tong University, Shanghai, 200240, China}
\affil[22]{School of Physics and Astronomy, Sun Yat-Sen University, Zhuhai, 519082, China}
\maketitle
\abstract{
We propose a major upgrade to the existing PandaX-4T experiment at the China Jinping Underground Laboratory. The new experiment, PandaX-xT, will be a multi-ten-tonne liquid xenon, ultra-low background, and  general-purpose observatory. The full-scaled PandaX-xT contains a 43-tonne liquid xenon active target. Such an experiment will significantly advance our fundamental understanding of particle physics and astrophysics. 
The sensitivity of dark matter direct detection will be improved by nearly two orders of magnitude compared to the current best limits, approaching the so-called ``neutrino  floor'' for a dark matter mass above 10 GeV/$c^2$, providing a key test to the Weakly Interacting Massive Particle paradigm. By searching for the neutrinoless double beta decay of $^{136}$Xe isotope in the detector, the effective Majorana neutrino mass can be measured to a [10 -- 41] meV/$c^2$ sensitivity, providing a key test to the Dirac/Majorana nature of neutrinos. Astrophysical neutrinos and other ultra-rare interactions can also be measured and searched for with an unprecedented background level, opening up new windows of discovery. Depending on the findings, PandaX-xT will seek the next stage upgrade utilizing isotopic separation of natural xenon. 
}



\section{Introduction}\label{sec:intro}
Dark matter (DM) and neutrinos are crucial matter ingredients in the Universe, both playing key roles in the evolution of the Universe and the formation of galactic structures. However, the nature of the dark matter remains unknown. Some fundamental properties of neutrinos are also elusive.
One of the leading candidates of DM is the so-called ``Weakly Interacting Massive Particles'' (WIMPs). They emerge naturally from various extensions of the Standard Model of particle physics, for example, Supersymmetry, and can reproduce the observed DM relic density without fine-tuning~\cite{Bertone:2004pz,Jungman:1995df}.
In 1985, inspired by the experimental development in searching for coherent neutrino-nucleus scatterings, Goodman and
Witten proposed to use similar detectors to search for rare collisions between DM particles and atomic nuclei~\cite{Goodman:1984dc}. This experimental approach, known as the DM direct detection, has been carried out around the globe for nearly forty years (see, for example, Refs.~\cite{Gaitskell:2004gd,MarrodanUndagoitia:2015veg,Liu:2017drf,Schumann:2019eaa,Billard:2021uyg,Cushman:2013zza} and references therein). In recent decades, the dual-phase xenon time projection chamber (TPC) has emerged as a leading experimental technique and made very fast progress on the detection sensitivity to DM-nucleon interactions, covering a wide range of parameter space for a DM mass greater than a few GeV/$c^2$. Three large experimental collaborations, PandaX~\cite{PandaX-4T:2021bab}, XENON~\cite{XENON:2023cxc}, and LZ~\cite{LZ:2022lsv}
have now all entered the multi-tonne-target stage. Yet, no clear DM signals have been observed. 
Ultimately, the sensitivity of DM direct detection will be limited by the neutrino background from the coherent scatterings of solar and atmospheric neutrinos with
nuclei, known as the ``neutrino floor''~\cite{Billard:2013qya} or ``neutrino fog''~\cite{OHare:2021utq}. It is worth noting that once the neutrino floor
is reached, the xenon detector is also a sensitive probe of neutrino interactions, which offers new physics opportunities on its own. 

The neutrino can be either different from or the same as its own anti-particle, known as the Dirac or Majorana fermion, respectively~\cite{Majorana:1937vz}. 
The Majorana nature of the neutrino would offer a natural explanation for the smallness of the neutrino mass via the so-called Seesaw mechanism~\cite{Minkowski:1977sc, yanagida1979proc, GellMann:1980vs, Glashow:1979nm, Mohapatra:1979ia}, and may also be 
connected to the matter-antimatter asymmetry in the Universe through Leptogenesis~\cite{Fukugita:1986hr}. It has been long realized that if neutrinos are Majorana fermions, an ultra-rare nuclear decay known as the Neutrinoless Double Beta Decay (NLDBD) will happen~\cite{Furry:1939qr}. NLDBD experiments can probe the effective Majorana neutrino mass ($m_{\beta\beta}$), a linear combination of three neutrino mass eigenvalues therefore dependent on the neutrino mass ordering (MO)~\cite{Agostini:2022zub}. 
Over the years, quite many NLDBD experiments have been carried out. 
The most recent ones search for NLDBD with the sensitivity of $m_{\beta\beta}$ at $\mathcal{O}(100)$~meV~\cite{GERDA:2020xhi,Majorana:2022udl,EXO-200:2019rkq,CUORE:2021mvw,SNO:2021xpa,NEMO-3:2015jgm,KamLAND-Zen:2022tow}.
KamLAND-Zen, which uses 800~kg of enriched $^{136}$Xe, has recently produced the best upper limit on $m_{\beta\beta}$ of [36 -- 156]~meV/$c^2$~\cite{KamLAND-Zen:2022tow}. Some future experiments, such as nEXO ($^{136}$Xe)~\cite{nEXO:2021ujk}, KamLAND2-ZEN ($^{136}$Xe)~\cite{Nakamura:2020szx}, LEGEND ($^{76}$Ge)~\cite{Brugnera:2023zgw}, CDEX-300$\nu$ ($^{76}$Ge)~\cite{Ma:2023yrk}, JUNO-DBD ($^{130}$Te/$^{136}$Xe)~\cite{Zhao:2016brs}, CUPID ($^{100}$Mo)~\cite{CUPID:2022jlk}, are aiming at a sensitivity of 15~meV/$c^2$ or better, to fully cover the parameter space corresponding to the so-called inverted neutrino MO. 

The DM direct detection and NLDBD experiments both pursue extremely low background and share similar detection techniques. However, due to vastly different energy regions and signal characteristics, and that NLDBD experiments normally require enriched isotopes and emphasize the energy resolution in the MeV region, the two types of experiments are mostly developed in parallel. Recently, securing large amounts of enriched isotopes for NLDBD searches has become a major challenge. On the other hand, XENON and LZ have demonstrated to achieve a sub-percent energy resolution around 2.5 MeV, the $Q$-value of $^{136}$Xe NLDBD~\cite{XENON:2020iwh,Pereira:2023rte}. Consequently, the approach of using a large natural xenon detector (containing $\sim$8.9\% $^{136}$Xe) to simultaneously pursue both avenues becomes increasingly attractive~\cite{Aalbers:2022dzr,DARWIN:2016hyl,XLZD}.

The China Jinping underground Laboratory (CJPL), located in Sichuan Province, China, with an overburden of 2,400~m, is currently the deepest underground laboratory in the world~\cite{Cheng:2017usi}. The extreme depth provides superb shielding to cosmic rays and related background, ideal for the experimental studies of DM and neutrinos. Two main experiments, the PandaX~\cite{PandaX:2014mem, PandaX:2018wtu} and CDEX~\cite{CDEX:2013kpt, CDEX:2014amu, CDEX:2018lau}, have been developing staged research programs on the DM and neutrinos at CJPL since 2009. The previous two generations of PandaX featuring a 120 kg (PandaX-I) and 580 kg (PandaX-II) liquid xenon (LXe) TPC, respectively, had produced competitive results in dark matter search~\cite{PandaX:2014ria,PandaX-II:2016andi,PandaX-II:2017hlx}, as well as in neutrino studies~\cite{PandaX-II:2020udv}.
PandaX-4T, with a sensitive target of 4 tonnes of LXe, is under operation in the newly expanded CJPL-II~\cite{Li:2014rca}. With only the commissioning data, it produced leading constraints to DM-nucleon and DM-electron interactions~\cite{PandaX-4T:2021bab, PandaX:2022xas, PandaX:2022xqx, PandaX:2023xgl} and the electromagnetic properties of the DM~\cite{PandaX:2023toi, BaiYangComment}, and made a precise determination of the two-neutrino double beta decay (DBD) lifetime of $^{136}$Xe~\cite{PandaX:2022kwg} and the best lower limits on the lifetimes of DBD and NLDBD of $^{134}$Xe~\cite{PandaX:2023ggs}. As CJPL-II is being transformed into a general-purpose National Major Scientific Facility in China, we propose a next-generation experiment, PandaX-xT, with a multi-ten-tonne active target,  strategically combining the DM and NLDBD efforts. When reaching its full scale, PandaX-xT is envisioned to contain 47 tonnes of natural xenon, of which 43 tonnes function as the active target. The main scientific objectives of PandaX-xT include, but are not limited to:
\begin{enumerate}
    \item Searching for DM-nucleon interactions with sensitivity at 90\% C.L. down to the neutrino floor, a key test on the WIMP DM paradigm;
    \item A stringent test on the Majorana nature of neutrinos using the NLDBD of $^{136}$Xe, covering most of the parameter space for inverted neutrino MO;
    \item Detecting low-energy neutrinos from the Sun and other astrophysical origins, and searching for ultra-rare new physics signals in general.
\end{enumerate}

In the rest of this paper, we shall discuss the conceptual design of PandaX-xT, the background projections in different energy regions, and its scientific potential.

\section{Conceptual Design of PandaX-xT}\label{sec:experiment}
\begin{figure}[hbt]
    \centering
    \includegraphics[width=0.45\textwidth]
    {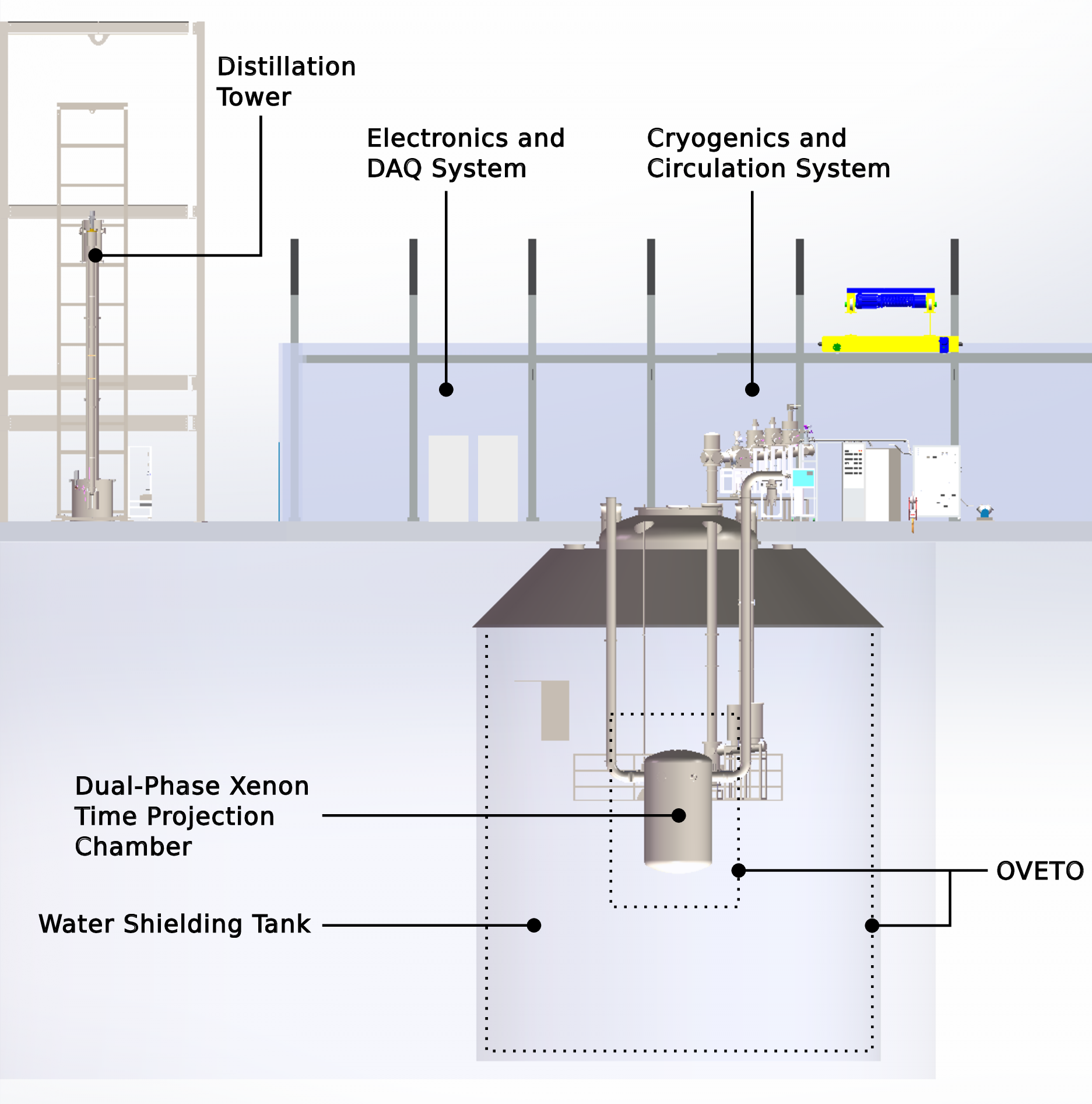}
    \caption{
    Schematics of the overall layout of PandaX-xT.}
    \label{fig:overall}
\end{figure}

\begin{figure}[hbt]
    \centering
    \includegraphics[width=0.6\textwidth]{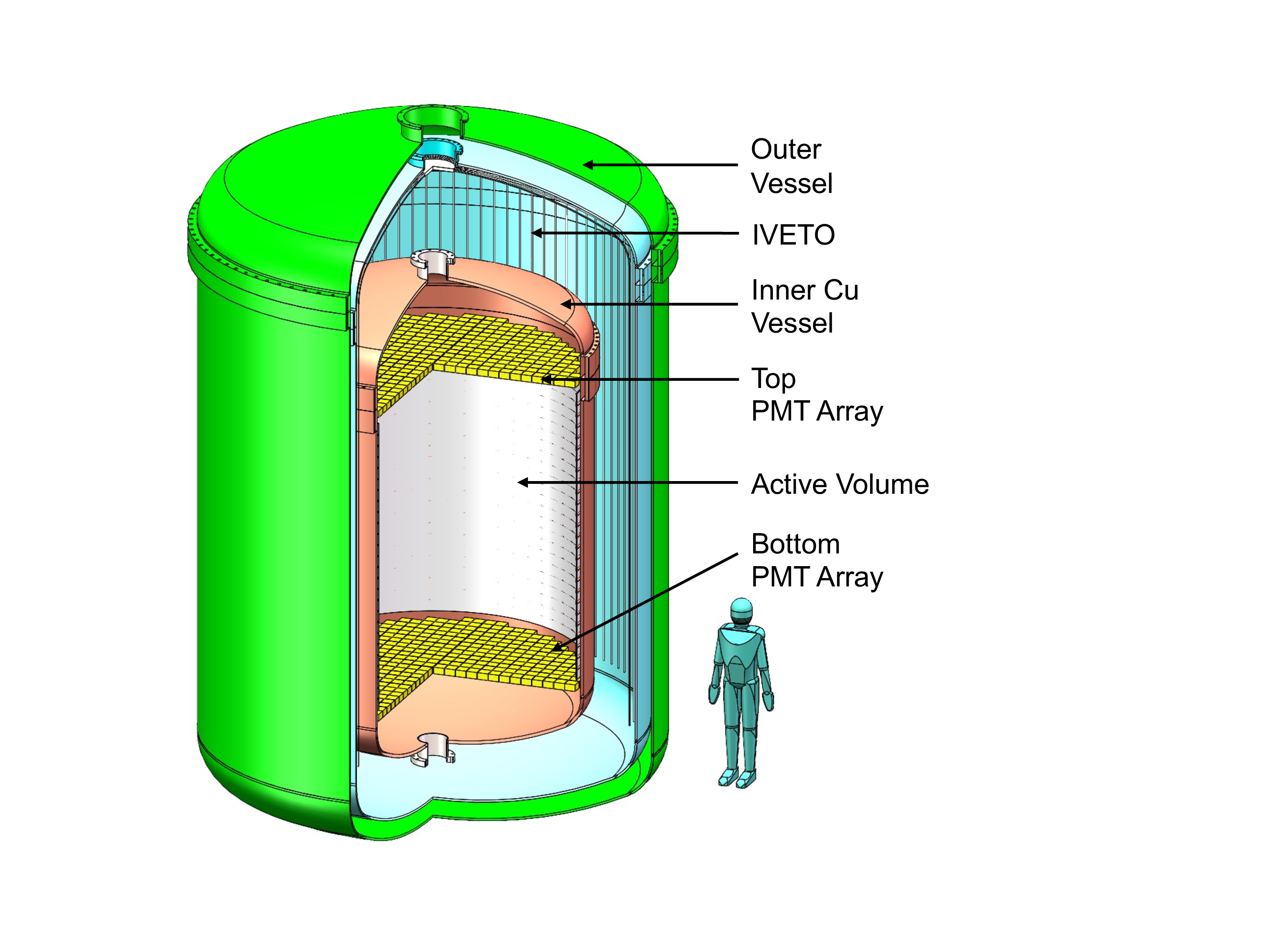}
    \caption{Schematic drawing of the conceptual design of the PandaX-xT cryostat and TPC. The IVETO region is 0.5~m thick between the IV and OV. Extra space above the IV is reserved for top feedthroughs and xenon tubes.}
    \label{fig:tpc_drawing}
\end{figure}

The overall design of the PandaX-xT is shown in Fig.~\ref{fig:overall}, reusing most of the existing lab infrastructure from PandaX-4T. 
The LXe detector is located in the center of an outer ultrapure water veto (OVETO). To achieve an extremely low background environment, the LXe cryostat in which the TPC is placed is made with a thin-walled ($\sim$3~mm) ultrapure copper inner vessel (IV), 
with a double-walled vacuum insulated low background titanium outer vessel (OV). In between the IV and OV, a low-temperature liquid scintillator (LS) volume is pressurized to balance the internal pressure of the IV, and also acts as an anti-coincidence detector to suppress the background (IVETO). The layout of the TPC, IVETO, and OV is shown in Fig.~\ref{fig:tpc_drawing}. To satisfy the needs of detecting DM, NLDBD, and astroparticle neutrinos, PandaX-xT is required to have an effective energy range between 100~eV and 10~MeV, an energy resolution of $\sigma$/E$<$1\% at 2.5~MeV, and a sub-cm vertex resolution.
The major detector components of PandaX-xT and key design features will be discussed below. 


The OVETO utilizes the passive water shielding in PandaX-4T, contained in a stainless steel (SS) tank with a total volume of 900~m$^3$ (13~m height and 10~m diameter). The radioactivity from the experimental hall is shielded by the water to a negligible level. The radioactivity of the water was measured during the PandaX-4T operation to be $\sim$10 $\mu$Bq/kg for $^{238}$U, $^{232}$Th, $^{40}$K and $^{222}$Rn, leading also to a negligible background contribution~\cite{PandaX-4T:2021lbm,Ma_2021}. 
We plan to instrument the water shielding with two layers of 8-inch PMTs, separated by Tyvek foil as optical reflectors.
The inner layer serves as a neutron veto with some gamma vetoing capability.
The inner PMTs will be mounted on the surface of a cylinder surrounding and looking toward the liquid xenon detector,
providing a photocathode coverage of $\sim$10\%. The neutrons are tagged by the 2.2~MeV gamma rays from hydrogen captures, and the tagging efficiency is estimated by a Geant4~\cite{GEANT4:2002zbu} simulation to be about 70\% assuming a trigger threshold of 5 fired PMTs.

At a later stage, to increase the light yield and thereby neutron tagging efficiency, we plan to blend about 5\% LS in the water using water-based liquid scintillator (WbLS).
The WbLS technology,
%
together with its potential applications, have been studied over a decade in the field \cite{Yeh:2011zz, Alonso:2014fwf, Bignell:2015oqa, Theia:2019non, Land:2020oiz, Theia:2022uyh}. 
The 5\% blending ratio is a modest assumption that is also used in the simulation of \textsc{Theia}, a proposed neutrino detector using WbLS \cite{Theia:2019non}.
We have also been experimenting different WbLS formulas for our application.

The outer layer consists of PMTs mounted on the wall of the cylindrical water container in a sparser arrangement compared to the inner layer, providing a very high efficiency in vetoing muons and muon-induced hadronic showers.
Additionally, with more PMTs, we are planning to further expand the OVETO from the 0.9~kton water tank to the water pit in which the water tank is situated.
The pit is going to be filled with 4.5~kton ultra-pure water, about a tenth of the mass of the Super-K detector.  Super-K detects about 10 atmospheric neutrino events per day above 100~MeV \cite{Suzuki:2019jby}.
By a simple scaling, we expect about one atmospheric neutrino event per day in the expanded OVETO, although the event rate will also be affected by other factors, such as the detection efficiency and the low latitude of the detector location \cite{Zhuang:2021rsg}.
Such in-situ measurement will provide data to potentially constrain the atmospheric neutrino modeling at the Jinping site.

The IVETO is a layer of 0.5~m thick LS instrumented with plastic wavelength shifting (WLS) fibers coupled to an array of photosensors.
The LS will be under the liquid xenon temperature, which is about $-100^\circ$C.
The neutron capture signals in LS or partial energy deposition from gamma rays from internal materials will be tagged to provide an effective veto for the otherwise candidate events happening in the TPC.
The design is still under R\&D with open outcome, but is based on several existing technologies:
\begin{itemize}
    \item 
    The cold liquid scintillator under $-50^\circ$C has been successfully developed for the TAO experiment\cite{JUNO:2020ijm, Xie:2020bqa}. The cocktail is a LAB-based LS mixed with PPO, bis-MSB and ethanol.
    We are exploring the LS performance under $-100^\circ$C, starting with the recipe used in TAO.
    \item 
    The use of the WLS fibers to collect and shift the wavelength of the scintillation photons 
    is a widely-used approach (see, e.g., Refs.~\cite{MINOS:2002xlc, NOvA:2007rmc}). 
    Importantly, the fibers can work stably under cryogenic temperature as shown in GERDA experiment where the fibers were immersed in liquid argon \cite{GERDA:2017ihb}.
    \item
    At the LXe temperature, without trying to remove the water content in the LS as in TAO, it was observed that the LAB-based LS stays as a thick translucent fluid. As already demonstrated by the LiquidO project~\cite{Buck:2019tsa}, opaque LS in fact
    helps the photon collection by the WLS fibers, due to localized scintillation photons as a result of a short scattering length. 
    \item
    The LS can provide pressure balance and additional shielding for the copper inner cryostat. The basic idea was pioneered by EXO-200, where cryogenic fluid (HFE-7000) is filled between the cryostat vessel and the LXe vessel~\cite{Auger:2012gs}. The pressure balance allows a significant reduction in the thickness of the copper cryostat to minimize the radioactive background.
    For PandaX-xT, we are adding the veto functionality by using the cold LS instead of cryogenic fluid.
\end{itemize}
Based on our optical simulation, assuming a typical light yield of 10,000 photons per MeV for the LS at LXe temperature, and a Mie scattering length (which determines the opaqueness of the LS) of 3 mm, a 200 keV energy deposition with the distance to fiber of 5 mm gives approximately 5 PEs at the readout with an avalanche photodiode.
Therefore, in this study, 200~keV is assumed as the threshold for the IVETO. 
The veto efficiency for low-energy neutron background events is estimated to be 
about 86\%
via a simulation (Sec.~\ref{sec:bkg}). The neutron background levels presented in later Table~\ref{tab:bkg_below_10keV} assume the application of IVETO, and they are small even without the IVETO. However, IVETO will be an important insurance if the actual levels of radioactivities are worse than the specifications (see Sec.~\ref{sec:bkg}). In addition, IVETO will serve as a pressure balancer for the thin-walled Cu vessel (allowing for reducing the vessel's material) and a thermal buffer for the TPC.


%

The cryostat system holds the liquid xenon target and TPC at 178~K, with a maximum working pressure of 0.35~MPa, while preventing contamination by external radioactivity and electronegative impurity. To meet the low background requirements of the PandaX detector, as mentioned earlier the IV is made with a thin-walled oxygen-free copper container with a diameter of $\sim$2.65~m and a total height of $\sim$4~m containing the liquid xenon. The copper vessel thickness is kept thin to minimize its mass and radioactivity. The OV is a double-walled vacuum-insulated titanium vessel with an outer (inner) diameter of $\sim$4~m (3.5~m) and an outer (inner) height of $\sim$6~m (5~m). In between the IV and OV is a 0.5~m-thick LS IVETO region, with its pressure balanced with the liquid xenon, so the actual pressure is held by the OV. Special care has to be taken in the copper vessel design due to the non-uniform pressure caused by the LXe height, as well as in the control of pressure during the filling and emptying of the detector. The LXe TPC defines the active volume of the detector. 
In the conceptual design, the PandaX-xT TPC is 2.55~meters in diameter and 2.95~meters in height. It is enclosed by large, photon-transparent electrodes (the cathode, gate, and anode), which are made from wires arranged linearly in a single direction or as a mesh. The corresponding detector active volume contains approximately 43~tonne of liquid xenon. 
Dual-phase xenon TPC technology will be employed as the baseline technology, in which the ionization signal is converted into proportional scintillation in the gaseous xenon between the gate and anode. Alternative options, for example, with proportional scintillation produced in liquid xenon~\cite{Aprile:2014ila, Ye:2014gga, Juyal:2019gch} will also be explored. Since xenon procurement will likely take some time, the volume of the TPC needs to be easily upgraded based on the amount of xenon in possession. An acrylic barrel will be used as the major structural holder of the TPC. There is a very good assurance on the low radioactivity~\cite{DEAP-3600:2017ker} since the acrylic panels used in JUNO have achieved a part-per-trillion (ppt) level of radio-impurity in mass production~\cite{JUNO:2021kxb, CAO2021165377}. 
The inner surface of the acrylic will be covered by a cylindrical sleeve of Polytetrafluoroethylene (PTFE) reflector of 2~mm-thickness to enhance the photon collection efficiency (VUV photon transmission for 2~mm PTFE is about 1\%~\cite{Althuser:2020uxc}). The field-shaping circuitry will be implemented on a cylindrical flexible printed circuit board sandwiched between the PTFE sleeve and acrylic barrel. 
The top and bottom of the TPC will be enclosed by two acrylic plates which contain holes to mount and support photosensors.
Such a design is sufficiently robust and straightforward to be fabricated. Assuming that the photosensor arrays are constructed into the final form, the acrylic barrel structure will be the only component that gets swapped for each upgrade.
One important design goal for PandaX-xT is to maximize the active volume. The volume of the 2-inch sqaure-shape photosensors (see later) is significantly smaller than the 3-inch cylinder-shape ones used in previous generations of PandaX, LZ, and XENON. Such photosensor choice, in combination with the low background acrylic fillers installed below the bottom array, will significantly reduce the dead volumes at the bottom of the detector. The xenon self-shielding areas in and around the TPC can be minimized given the reduction of material radioactivities and the IVETO/OVETO systems outside the inner vessel. With these measures, we target to control the dead volume to below 10\% of total xenon ($\sim$47~tonne).
The basic TPC performance parameters are: 1) a photon detection efficiency ($g_1$) of $>$10\%, 2) a drift field of $>$100~V/cm to allow for a 99.7\% rejection of electron-recoil (ER) events, while maintaining a 50\% efficiency on the DM-nuclear recoil (NR) events. These specifications have been mostly achieved in PandaX-II and PandaX-4T~\cite{PandaX-II:2016andi,PandaX-4T:2021bab}. The performance will be ensured through intensive calibrations, including the injection of radioisotopes to determine detector optical parameters, uniformity, and electron recoil (ER) responses~\cite{Kastens:2009pa,LUX:2015amk,XENON:2016rze,Ma:2020kll}. Additionally, various neutron sources, such as the deuteron-deuteron generator, will be used to obtain nuclear recoil (NR) responses~\cite{akerib2016low,aprile2019xenon1t}.


The photosensor arrays collect the 178~nm photons produced in the liquid and gaseous xenon, providing energy measurement
as well as locations and sequences of energy depositions. The photosensitive area of PandaX-xT will exceed 10~m$^2$. 
In the past, significant saturation effects were observed in the Hamamatsu R11410 3-inch photomultiplier tubes (PMTs) in the MeV regime~\cite{XENON:2020iwh,PandaX:2022kwg,Luo:2023ebw}, limiting the energy linearity and resolution, and impairing the ability to reconstruct scattering positions in the LXe TPCs. Compared to R11410, Silicon Photomultiplers (SiPMs) have much better granularity and energy linearity, but unfortunately with relatively higher dark count rates under the LXe temperature (best-reported value 0.035-0.069 Hz/mm$^2$~\cite{Sakamoto:2023ond}, 3-7 times higher than that of R11410~\cite{Barrow:2016doe}). The implementation of SiPM electronics would also be challenging due to the large number of channels. Therefore, a new type of photosensor with a square area of about 2.5$\times$2.5 cm$^2$ per channel (Hamamatsu R12699) are being considered as the primary option for PandaX-xT, while we continue to closely monitor the technological advancements of SiPMs. In total, we expect a total of $\sim$8000 independent channels if both top and bottom arrays are instrumented with R12699. 
Their fast ($\sim$1.2ns) timing characteristics (vs. $\sim$5.5 ns for R11410) will provide more information on the scintillation waveforms. 

    
The readout electronics and data acquisition system digitizes the waveforms from photosensors and saves them on disk for future processing. They must support different data-taking rates in physics runs ($\sim$10~Hz) and calibration runs ($>$100~Hz). The readout system of PandaX-xT becomes much more challenging due to the large number of channels compared to PandaX-4T. The baseline solution involves using 500 MS/s 14-bit customized digitizers~\cite{PandaX_FADC_2021} which allows both single-channel-over-threshold readout or a firmware-based global readout trigger. Despite the maturity of the technology, it still needs to address the radioactivity and outgassing of thousands of mini-coaxial cables and the development of high-density low-temperature feedthroughs. As an alternative solution, cold electronics are under consideration for PandaX-xT. The system would be positioned near the photosensors inside the inner cryostat, digitize electrical signals and subsequently convert the data into optical signals using low-power transceivers. The system should feature high-speed sampling ($>$ 100 MS/s) and low-power consumption ($<$100~mW/channel). Additionally, the front-end electronics should function fully at 178 K temperature and exhibit low radioactivity. The increasing sampling rate and expanding number of readout channels present significant challenges in data processing and storage. During the standard background data taking, assuming the sampling rate of 500 MS/s, it is estimated that over 5~TB of raw data will be generated per day. An upgrade of the data reduction strategy developed for PandaX-4T has the potential to effectively address these challenges~\cite{Zhou:2023vmz}.
    
The LXe handling system for the filling, recovery, and storage of xenon, dubbed First-X, allows xenon transfer in and out of the PandaX-xT detector entirely under the liquid form~\cite{Wang:2023wrr,Distillation:2017}. This is imperative since the traditional filling and recuperation process with gaseous xenon becomes too time-consuming. The First-X system consists of multiple storage tanks, two cryogenic pumps, and the liquid nitrogen (LN2) and gas nitrogen (GN2) storage modules. Each storage tank is designed to store over 6~tonnes of xenon with an inner volume of 6.3~m$^{3}$ and a design pressure of 7.1~MPa, satisfying the needs for long-term xenon storage in liquid phase (178~K, 0.2~MPa) or gaseous phase (300~K, 7.0~MPa). The storage tank's pressure can be rapidly regulated by three heat exchangers including two independent LN2-based cooling modules and one GN2-based heater. The cryogenic pump is a centrifugal pump with a magnetic drive that isolates the pump and LXe to avoid radioactivity contamination from the electric motor~\cite{Barber}. The First-X system with the initial six storage tanks has been constructed. The LXe filling/recovery mass flow rate has been demonstrated to reach 2.1~tonnes/hour. 

The cryogenics and recirculation system provides cooling power for the LXe, while removing electronegative impurities with an active circulation loop and a purifier. The concept design is based on the successful operation of the cooling buses used in previous generations of PandaX~\cite{Gong:2012thh,Zhao:2020vxh}.
A preliminary estimate of the total heat load for PandaX-xT is about 1300~W at 178~K, including the radiation and conduction thermal loss of the IV (400~W), heat load of the gaseous circulation (100~W), and heat production by the frontend electronics ($\sim$800~W, if the option of cold electronics is adopted).
A customized cooling tower with at least two GM Cryocoolers (Cryomech AL600, each with $\sim$900~W cooling power at 178~K) will provide needed cooling power during regular operation.
At the beginning of the experiment, an additional 2.5~kW of assistant cooling power~\cite{ThermalManagement} will be needed to cool down the LS as well as the IV and OV to 178~K in 20~days.
For this need, alternative means of cooling such as liquid nitrogen~\cite{Giboni:2019fqo} will be incorporated via parallel cooling heads and loops. 
In addition, to achieve a long electron lifetime in the drifting field ($>$2 ms, corresponding to an electronegative impurity level of $10^{-10}$~mol/mol), a sufficiently fast recirculation speed is needed on both the liquid and gaseous xenon. The traditional gas recirculation with hot zirconium purifiers and heat exchangers can be used with a customized magnetically-coupled piston gas pump~\cite{Piston1,Piston2}. A few such loops operating in parallel can achieve a 500~standard liters per minute (slpm) flow rate. Moreover, since O$_{2}$ is usually the dominant contributor to the attenuation of ionized electrons in liquid xenon, the liquid xenon recirculation ($\sim$2~lpm) with a cryogenic pump and O$_{2}$ sorbent will be used in parallel. This approach has been demonstrated to work very effectively in XENONnT~\cite{XENON:2022ltv,Liquid_purity:2022}. Special measures of radon control will be taken to avoid introducing radon into the TPC through the recirculation loops. 

The distillation system removes the radioactive impurities such as $^{85}$Kr, $^{222}$Rn and tritium from the xenon. Krypton and radon are noble gas impurities originating either from external leaks or internal emanations, which cannot be removed by commercial purifiers. Tritium background was introduced in PandaX-II and PandaX-4T through tritiated-methane calibration injection, and the removal efficiency from the commercial purifier also appeared insufficient for a tritium level below $10^{-24}$ mol/mol~\cite{PandaX-II:2020udv,PandaX-4T:2021bab}. Cryogenic distillation is an effective technique to achieve ultra-pure xenon with a large flow rate based on the different boiling points of Xe, Kr, Rn, and tritiated methane (165 K, 120 K, 211 K and 111.5 K at atmospheric pressure, respectively), as demonstrated in earlier generations of PandaX~\cite{Distillation:JINST2014,Distillation:JINST2021}. The requirements from PandaX-xT are more stringent  (see Sec.~\ref{sec:bkg}), with levels of $^{85}$Kr, $^{222}$Rn and tritium lower than 2~nBq/kg, 0.5~$\mu$Bq/kg, and 0.01~nBq/kg, respectively. These are challenging specifications, and there have been a number of completed and ongoing developments by PandaX~\cite{Wang:2023wrr,Distillation:JINST2014,Distillation:JINST2021,Distillation:JINST2021-2,Distillation:JINST2023}. Quite some developments are also reported by XENON collaboration in this area~\cite{Distillation:2017,Distillation:2012}.
A new krypton-removal tower and radon-removal tower are under construction. According to the design, the krypton content in xenon could be reduced to 1$\times 10^{-14}$ mol/mol (2 nBq/kg) at the flow rate of 30 kg/h with 99 $\%$ xenon collection efficiency. The krypton tower is expected to remove tritium in the xenon to a completely negligible level.
For the radon tower, with a 5~lpm liquid xenon flow rate (860 kg/h), the radon reduction can reach a factor of 10 lower~\cite{Distillation:JINST2023}, achieving a level of 0.5 $\mu$Bq/kg compared to 5 $\mu$Bq/kg level in PandaX-4T without radon distillation~\cite{PandaX-4T:2021bab}. These, together with other means of strict radioactivity control, are expected to lower the internal radioactive gas content to the desired values.

\section{Background Estimation}\label{sec:bkg}

The background level is the major driver of the scientific potential of PandaX-xT. For a multi-ten-tonne natural LXe detector, there are three major classes of background, 1) background arising from detector materials, 2) internal background from $^{85}$Kr, $^{222}$Rn, and tritium, and 3) physical background such as DBD events from $^{136}$Xe and neutrino scattering events. 
The first component exhibits strong position dependence at the outskirts of the detector, so a proper fiducial volume selection can significantly reduce its contribution.
The latter two components are uniformly distributed inside the detector. 
In this study, we assume that the background from the detector materials is primarily due to photosensor arrays and the copper vessel. The radioactivity levels are taken to be 0.02, 0.01 and 1.5~mBq/cm$^2$ for photosensors for $^{238}$U, $^{232}$Th, and $^{40}$K, respectively, based on prototype sample measurements, and 5~ppt of $^{238}$U, $^{232}$Th, and $^{40}$K for the copper vessel (achieved in Majorana Demonstrator, EXO-200, and SuperCDMS experiments~\cite{MAJORANA:2016lsk, PhysRevC.92.015503, PhysRevD.95.082002}). The radioactivity of PTFE in the TPC is assumed to be 1 ppt for $^{238}$U and 10 ppt for $^{232}$Th, producing a negligible background. For the internal background in LXe, we assume a level of 2~nBq/kg and 0.5~$\mu$Bq/kg for $^{85}$Kr and $^{222}$Rn, respectively. For the physical background, the rates and energy spectra are taken from Ref.~\cite{Billard:2013qya}, except those for the double electron-capture of $^{124}$Xe, which are taken from the recent measurement from XENONnT~\cite{XENON:2022evz}. The cosmogenically produced $^{137}$Xe is assumed to be two orders of magnitude less in comparison to that estimated in DARWIN~\cite{DARWIN:2020jme}, scaled using the total muon rate ratio between the CJPL and LNGS~\cite{JNE:2020bwn}.
 
A customized Geant4 simulation, BambooMC~\cite{Chen:2021asx}, is used to estimate the background in PandaX-xT using a simplified geometry in Fig.~\ref{fig:tpc_drawing}. The approach is very similar to that for Pandax-4T~\cite{PandaX:2018wtu,PandaX-4T:2021lbm}. The intrinsic decay class in Geant4 is used to simulate radioactive decay chains, with all energy depositions inside the detector and veto detectors recorded, allowing a straightforward evaluation of the ER background. The neutron production arising from material radioactivities is simulated using a neutron generator improved from the traditional SOURCES-4A~\cite{osti_15215}, which utilizes an updated nuclear cascade model and fission model implemented in Geant4 to properly take into account correlated emissions of neutron and gamma ray(s) as well as multiple neutrons~\cite{PandaX-II:2019jmf, Chen:2021asx}.

Four sets of cuts are developed in simulation to select candidates, namely the single-scatter cut, veto cuts, energy region of interest (ROI) cuts, and fiducial volume (FV) cuts:
\begin{enumerate}
\item Single-scatter cut rejects events with multiple energy depositions in the detector simulation. 
\item Veto cuts remove events with energy deposition greater than 200 keV in the IVETO region. While the OVETO will operate at the same time providing additional veto efficiency, we omit its contribution in this study for conservativeness.
\item ROI cuts are defined for three main physics regions. The DM ROI, [1, 10]~keV$\rm{}_{ee}$, includes both low energy ER and NR background (neutron and neutrino-nuclear coherent scattering) events. The solar neutrino ROI, [10, 150]~keV$\rm{_{ee}}$, refers to the primary energy region for elastic-electron scattering signals from solar $pp$ neutrinos~\cite{PandaX:2024jjs}. The NLDBD ROI, [2433, 2483]~keV$\rm{_{ee}}$, covers a 50-keV window around the $Q$-value of $^{136}$Xe NLDBD (2458 keV). To implement a simplified energy response model, the ER energy deposition is smeared with a resolution function $\sigma/E$ of $30\%/\sqrt{E/{\rm keV}}+ 0.4\%$.
The NR energy is converted into an equivalent ER energy using the Lindhard factor~\cite{MSzydagis_2011, szydagis2018noble} but without the smearing.
\item FV cuts are optimized with a simple ``cut-and-count'' figure-of-merit $\displaystyle{\frac{\epsilon M\times T}{\sqrt{B}}}$ in which $\epsilon$ is the signal efficiency, $M$ is the fiducial mass, $T$ is the data taking time, and $B$ is the corresponding number of background events. For the DM ROI, since the signal is the DM-nucleus NR scattering, we assume an ER background rejection power of 99.7\% and an NR acceptance of 50\% with the standard ionization-to-scintillation ratio cut in LXe DM experiments.
There is only ER background in the solar neutrino and NLDBD ROIs. The resulting optimized cylindrical FV for the DM and solar neutrino region is 34.2 tonne (190 mm to the top, 170 mm to the bottom, and 102 mm to the side of the TPC) and NLDBD is 8.4 tonne (780 mm to the top, 816 mm to the bottom, and 484 mm to the side of the TPC) of natural xenon.
\end{enumerate}

\begin{figure}[hbt]
    \centering
    \includegraphics[width=0.45\textwidth]{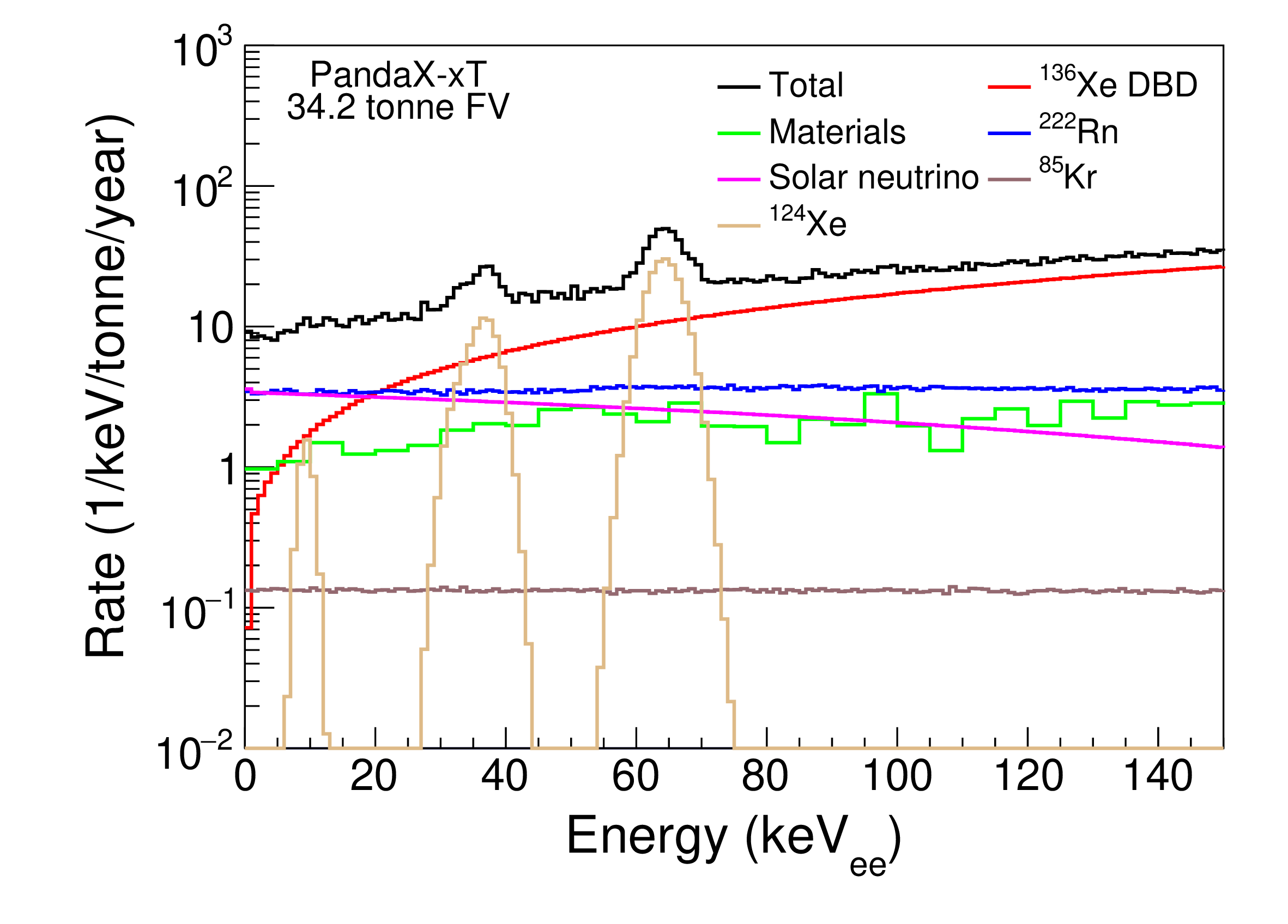}
    \caption{The energy spectra of individual background components between 1 and 150 keV (DM and solar neutrino regions) with SS, veto, and FV cuts applied.  }
    \label{fig:DMbkg}
\end{figure}

\begin{figure}[hbt]
    \centering
    \includegraphics[width=0.45\textwidth]{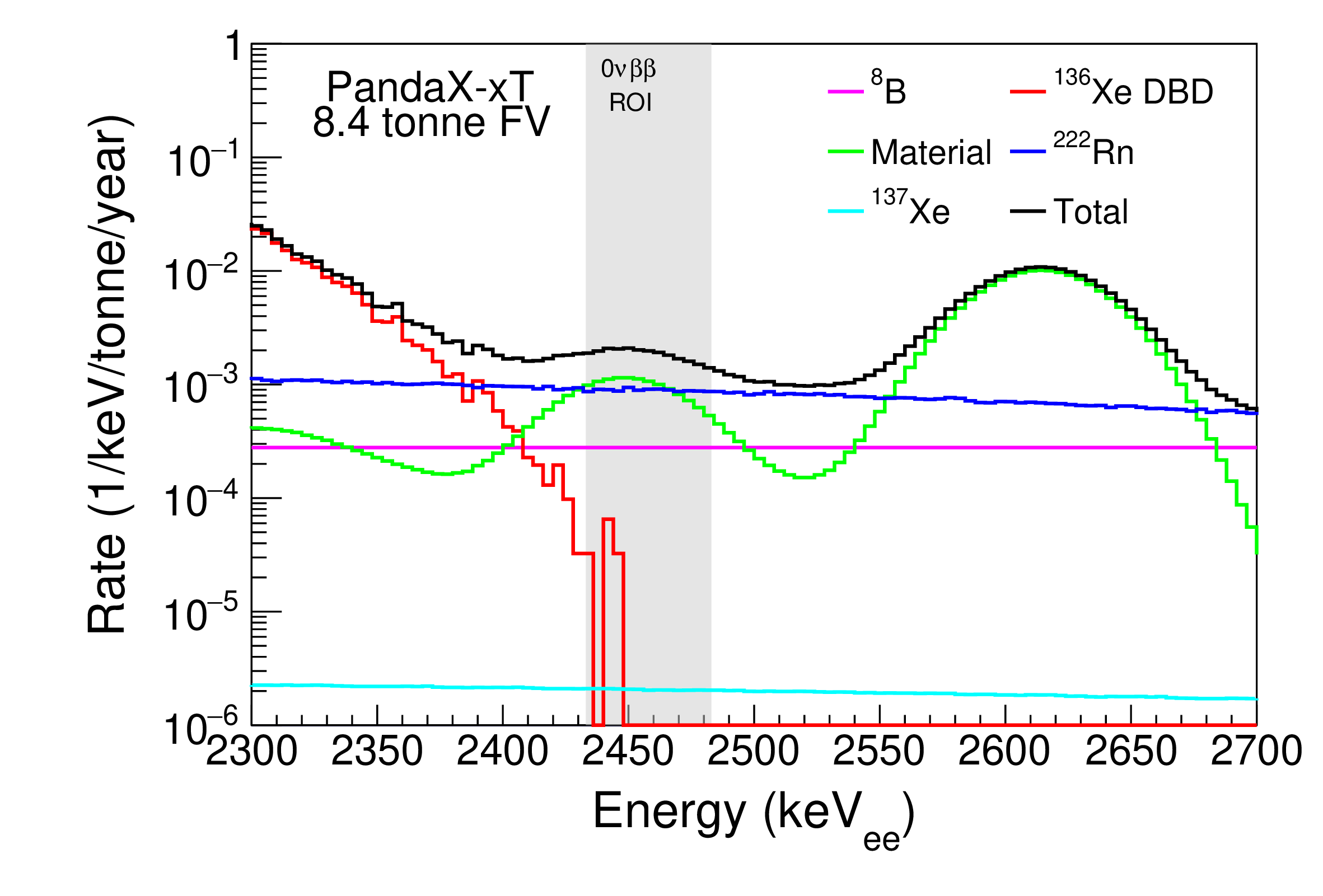}
    \caption{
    The energy spectra of individual background components in the MeV region with SS, veto, and FV cuts applied under the ``baseline'' background assumption. See text for details.}
    \label{fig:0vbbbkg}
\end{figure}

After these cuts, background energy spectra below 150 keV (the DM and solar neutrino-electron scattering
regions) and the high energy NLDBD region are shown in Figs.~\ref{fig:DMbkg} and \ref{fig:0vbbbkg}. Corresponding background events in the DM, solar neutrino, and NLDBD energy regions are summarized in Tables~\ref{tab:bkg_below_10keV}, ~\ref{tab:solar_nu_background}, and ~\ref{tab:nldbd_background}. Note that the background rate in the NLDBD ROI is highly sensitive to the background assumption, so the results of a ``baseline'' and ``ideal'' assumption are given. Under the 
``baseline'' assumption, one major  background arises from the $^{214}$Bi $\beta$-decay, the daughters of the internal $^{222}$Rn daughter. Although $^{214}$Bi can be very efficiently tagged and vetoed via the delayed coincidence of $^{214}$Bi--$^{214}$Po decay, we assume a 0.1\% residual untagged $^{214}$Bi nevertheless.
Under the ``ideal'' assumption, we assume that the material background is reduced by 5 times from the ``baseline'' and the background from $^{222}$Rn is negligible.

\begin{table}[hbt]
    \footnotesize
    \centering
    \begin{threeparttable}
    \caption{Expected background for the DM search region between 1 to 10 keV$\rm{}_{ee}$. The FV mass is 34.2 tonnes. The rates are obtained without applying the ER/NR rejection cut.}
    \label{tab:bkg_below_10keV}
    \doublerulesep 0.1pt 
    \tabcolsep 2pt 
    \begin{tabular}{l|cc} 
    \toprule
     & ER (event/tonne/year)  & NR (event/tonne/year) \\\hline
    Photosensors    & 8.3       & 0.00007        \\
    Copper vessel   & 0.02       & 0.000002        \\ 
    $^{85}$Kr       & 1.2       &   -           \\
    $^{222}$Rn      & 28.2      &   -          \\
    $^{136}$Xe      & 9.6       &   -           \\
    Solar $\nu$     & 27.5      &   -             \\ 
    Atmospheric $\nu$ & 0.0 & 0.02 \\
    Diffusive supernova $\nu$ & 0.0 & 0.002\\
    \hline
    \textbf{Total} & \textbf{74.8} & \textbf{0.02} \\
    \bottomrule 
    \end{tabular}
    \end{threeparttable}
\end{table}

\begin{table}[hbt]
    \footnotesize
    \centering
    \begin{threeparttable}
    \caption{Expected background for the solar neutrino-electron elastic scattering region between 10 to 150 keV for the 34.2-tonne FV. For comparison, the solar neutrino signal is calculated based on the Standard Solar Model~\cite{Vinyoles:2016djt}.}
    \label{tab:solar_nu_background}
    \doublerulesep 0.1pt 
    \tabcolsep 15pt 
    \begin{tabular}{l|c} 
    \toprule
     & ER (event/tonne/year)  \\
     \hline
    Photosensors & 270         \\
    Copper vessel & 21    \\ 
    $^{85}$Kr & 19   \\
    $^{222}$Rn & 439  \\ 
    $^{136}$Xe & 1927 \\ 
    $^{124}$Xe & 276 \\ 
    \hline
    Solar $\nu$  & 343 \\
    \bottomrule
    \end{tabular}
    \end{threeparttable}
\end{table}

\begin{table}[hbt]
    \footnotesize
    \centering
    \begin{threeparttable}
    \caption{Expected background under the baseline and ideal scenarios for the NLDBD energy range between 2433 and 2483 keV for the 8.4-tonne FV. See text for details.}
    \label{tab:nldbd_background}
    \doublerulesep 0.1pt 
    \tabcolsep 2pt 
    \begin{tabular}{l|cc} 
    \hline\hline
     & Baseline (event/tonne/year) & Ideal (event/tonne/year) \\
     \hline\hline
    Photosensors    & 1.4$\times$10$^{-2}$  & 2.8$\times$10$^{-3}$    \\
    Copper vessel   & 2.9$\times$10$^{-2}$  & 5.8$\times$10$^{-3}$     \\ 
    $^{222}$Rn      & 4.5$\times$10$^{-2}$  &  -   \\ 
    $^{136}$Xe DBD  & 5.2$\times$10$^{-4}$  &  5.2$\times$10$^{-4}$   \\ 
    $^{137}$Xe      & 1.0$\times$10$^{-4}$  &  1.0$\times$10$^{-4}$   \\ %
    Solar $^{8}$B $\nu$ & 1.4$\times$10$^{-2}$ & 1.4$\times$10$^{-2}$ \\ 
    \hline \hline  
    \textbf{Total}  &\textbf{1.1$\times$10$^{-1}$} & \textbf{2.4$\times$10$^{-2}$} \\
    \hline\hline
    \end{tabular}
    \end{threeparttable}
\end{table}

\section{Physics potential}\label{sec:sensitivity}
Being an unique deep underground and ultralow background liquid xenon observatory, the physics potential of PandaX-xT is very rich. Instead of presenting an exhaustive list of physics~\cite{aalbers2022next}, we benchmark the capability of PandaX-xT in the WIMP search, NLDBD, and solar neutrino detection, covering an effective energy range from keV to MeV.
Within these energy windows, the sensitivity of alternative signals with a single-scattering nature can be estimated based on the background rate presented in the previous section. The sensitivities to more exotic signals, for example, with multiple energy depositions via nuclear or atomic excitations or with time correlations, have to be studied case by case and are beyond the scope of this paper.

\begin{figure*}[bt]
    \centering
    \includegraphics[scale=0.5]{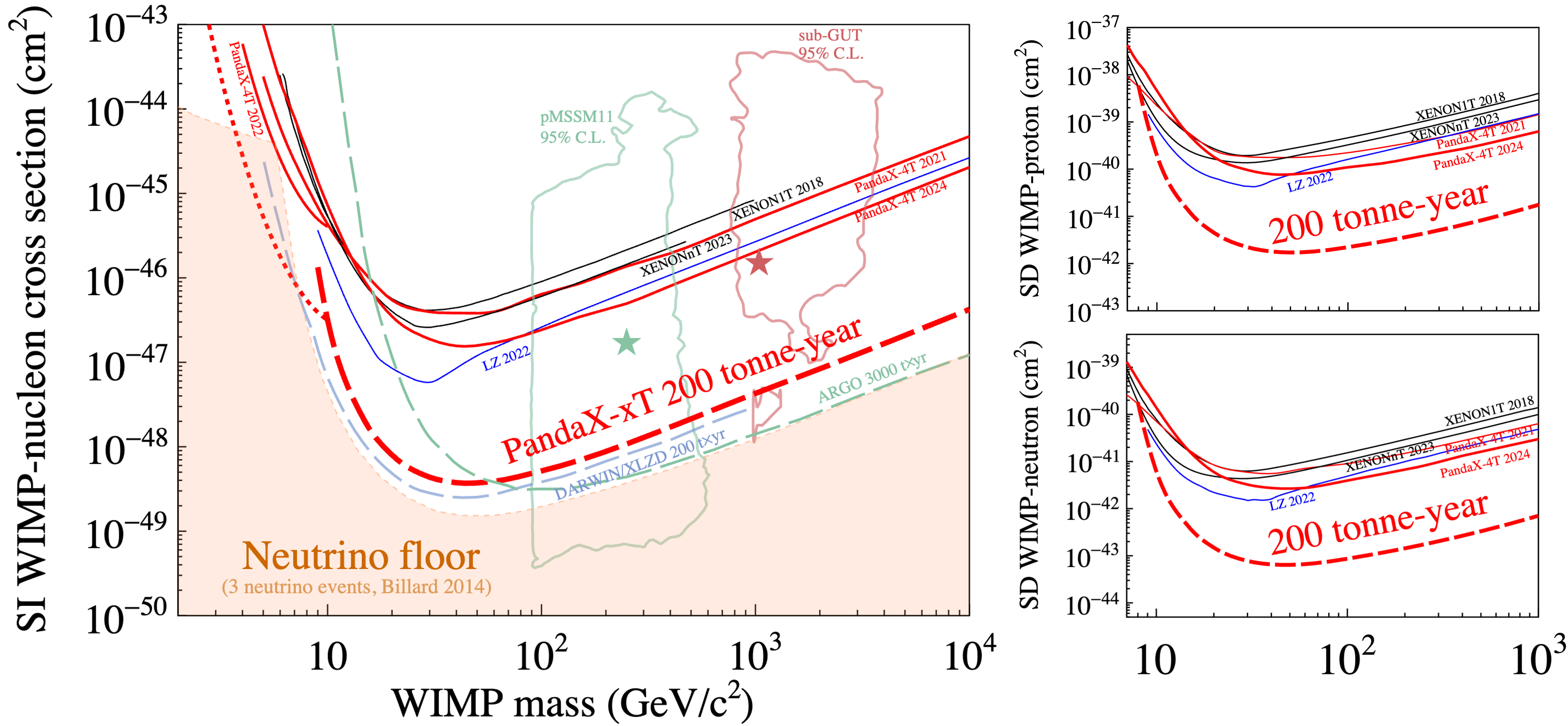}
    \caption{
    Projected 90\% C.L. exclusion sensitivities (red dashed curve) of PandaX-xT to spin-independent, and spin-dependent WIMP-neutron/-proton interactions estimated with a cut-and-count method, assuming the background level from Table~\ref{tab:bkg_below_10keV}, a 99.7\% ER rejection power and a 50\% NR acceptance, and an exposure of 200\,tonne$\cdot$year.
    Our benchmark energy range from 1 to 10 keV$\rm_{ee}$ has limited our sensitivity to DM masses less than 10\,GeV/$c^2$.
    The sensitivity curve for low-mass DM below 10\,GeV/$c^2$ (dotted line in the left graph) is estimated based on 
    the PandaX-4T efficiency curve in a specialized low energy analysis in Ref.~\cite{PandaX:2022aac}, and a 1:1 ratio of $^8$B neutrino induced CE$\nu$NS and accidental coincidence background. 
    The ``neutrino floor'' (3 events) and 95\% C.L. contour of allowed parameter space in Supersymmetry models are taken from Refs.~\cite{Billard:2013qya} and~\cite{Bagnaschi:2018}, respectively.
    Past experimental constraints~\cite{PandaX-4T:2021bab, PandaX-II:2017hlx, PandaX-II:2016andi, XENON1T2018SI, LUX2017SI, PandaX:2022xas, xenon1t_sd_2019, lux_sd_2017, XENON:2023cxc, LZ:2022lsv} are also overlaid.
    We also overlay sensitivities of the future DARWIN/XLZD~\cite{Baudis:2024_XLZD} and ARGO~\cite{McDonald:2024osu} with exposure indicated.
    }
    \label{fig:sisd}
\end{figure*}

For the WIMP search in the DM ROI, the background level is 75 events/tonne/year for electron recoil (ER) and 0.02 events/tonne/year for NR, respectively (Table~\ref{tab:bkg_below_10keV}). 
With the simple ``cut-and-count'' assumptions of an ER background rejection power of 99.7\% and an NR acceptance of 50\%, the ER and NR backgrounds in the ROI are 0.23 events/tonne/year and 0.01 events/tonne/year, respectively. Note that the accidental background has been omitted in the WIMP ROI to be consistent with the assumptions in previous studies, e.g. by DARWIN~\cite{Macolino:2020uqq}, and they are sub-dominant in the current generation experiments. 
With the standard DM halo and nuclear form factors parameters, and the spin-independent or spin-dependent isospin-conserving DM-nucleon interaction~\cite{Baxter:2021pqo}, and with a projected exposure of 200\,tonne$\cdot$year, the sensitivities are summarized in Fig.~\ref{fig:sisd}. Note that the PandaX-xT sensitivity reached about 3.5$\times$10$^{-49}$\,cm$^2$ for spin-independent interactions at a DM mass of 40 GeV/$c^2$. The sensitivity is expected to improve with a likelihood approach with the full distributions of the signal and background incorporated. 
Nevertheless, the limit covers most of the allowed parameter space from the constrained minimal Supersymmetry models~\cite{Bagnaschi:2018}. Also overlaid is the neutrino-background-only sensitivity ``floor'' with three neutrino-nucleus coherent scattering events from Ref.~\cite{Billard:2013qya}. The sensitivities to the spin-dependent interactions with the proton and neutron are also given in the figure.

\begin{figure*}[hbt]
    \centering
    \includegraphics[width=.8\textwidth]{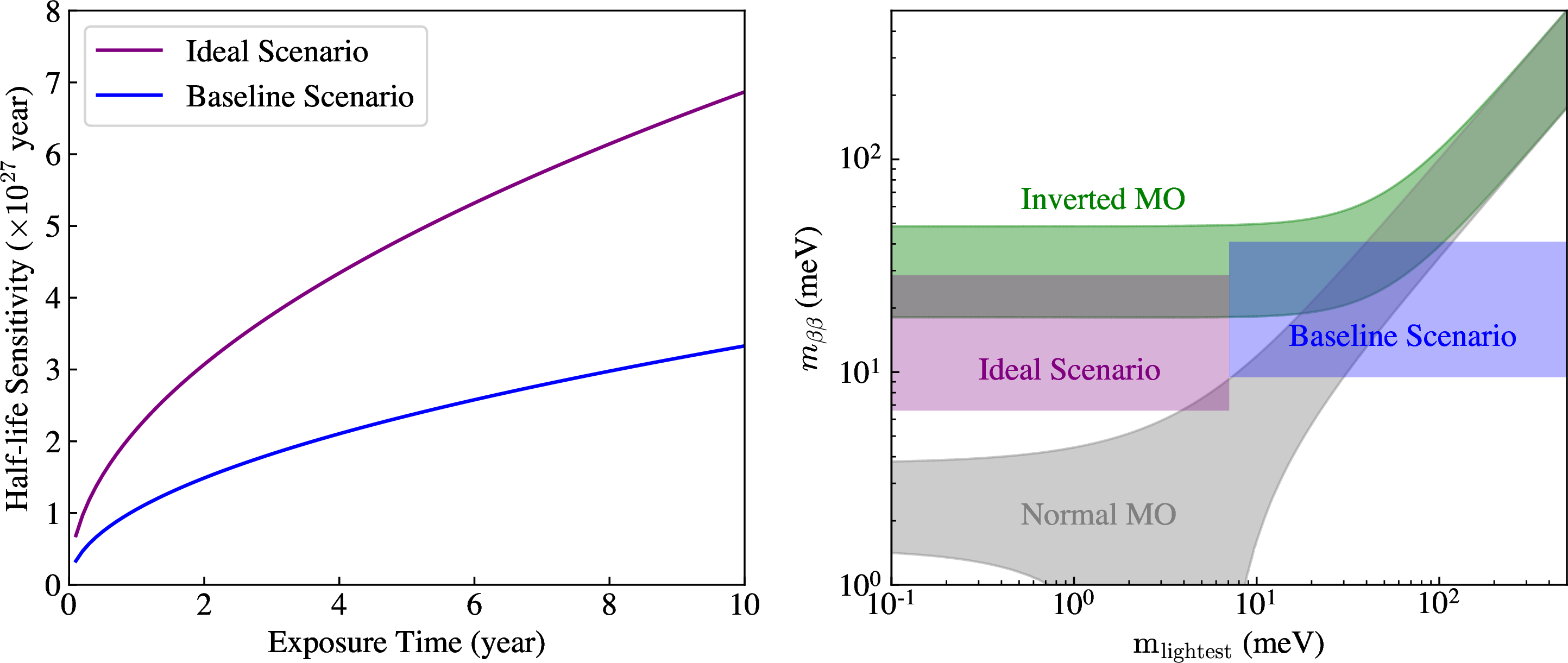}
    \caption{(Left) Projected 90\% C.L. sensitivity to NLDBD half-life as a function of exposure time for the baseline and ideal background scenarios, assuming a fiducial mass of 8.4-ton natural xenon. Inputs used in the calculation are in Table~\ref{tab:nldbd_background} and discussed in the text. (Right) Effective Majorana mass ($m_{\beta\beta}$) sensitivity with 10 year of exposure with respect to the phase space of the inverted and normal MO of neutrinos. The sensitivity bands of the two scenarios are due to the spread of commonly-used nuclear matrix elements.
    The two sensitivity bands are staggered intentionally for better visibility.
    }
    \label{fig:nldbd}
\end{figure*}

For NLDBD, at the $Q$-value of $^{136}$Xe NLDBD (2458 keV), and with an energy resolution $\sigma/E$ of 1\%, the level of the background for the 8.4-tonne fiducial volume 
is 2.1$\times 10^{-3}$ count/keV/tonne/year within a 50~keV energy window (Table~\ref{tab:nldbd_background}).  
The sensitivity to the NLDBD half-life can be expressed as
\begin{equation}
T^{\rm NLDBD}_{1/2} = \displaystyle{\frac{\ln 2 \times \epsilon\times f_{\rm ROI} \times (\text{\# of $^{136}$Xe atoms}) \times T}{1.64\sqrt{B}}},
\end{equation}
in which $f_{\rm ROI}= 0.68$ is the fraction of events enclosed in the NLDBD ROI, $\epsilon$ is the signal selection efficiency which is assumed to be 100\%, $B$ is the number of background events (Table~\ref{tab:nldbd_background}), the number of $^{136}$Xe atoms is calculated assuming natural abundance, and $T=$10~years. This leads to a half-life sensitivity of $3.3\times10^{27}$ and $6.9\times10^{27}$ years for the baseline and ideal background assumptions, respectively. The baseline result corresponds to an upper limit of the effective Majorana mass between [10 -- 41]~meV/$c^2$ on $m_{\beta\beta}$, in which the spread is due to the uncertainty in the nuclear matrix elements~\cite{Dolinski:2019nrj}. The sensitivity can mostly cover the allowed space for inverted neutrino MO, assuming that NLDBD is mediated by light Majorana neutrinos, as illustrated in Fig.~\ref{fig:nldbd}.

\begin{figure*}[hbt]
    \centering
    \includegraphics[width=0.6\textwidth]{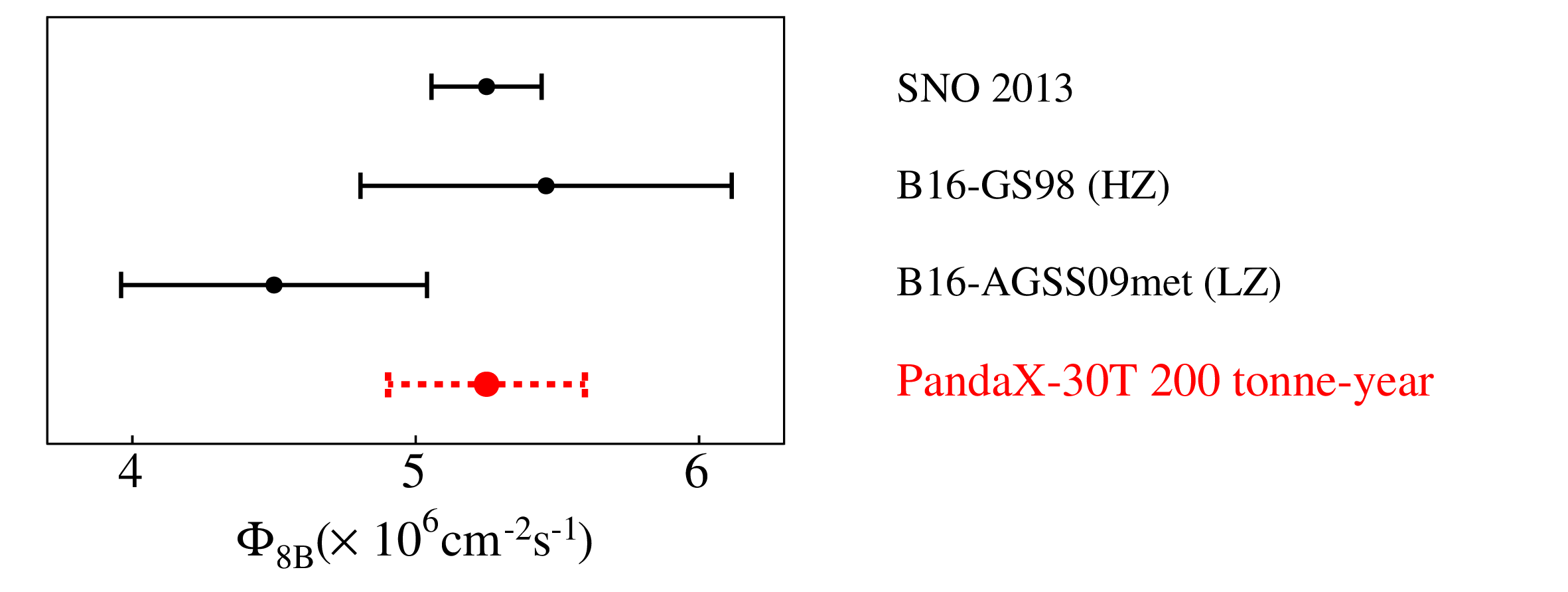}
    \caption{Projected sensitivity of PandaX-xT to solar $^8$B neutrino flux assuming the same mean value as SNO~\cite{aharmim2013combined}, compared with theoretical models~\cite{vinyoles2017new}.
    }
    \label{fig:b8_flux}
\end{figure*}

For the solar neutrinos, we benchmark our sensitivity to the solar $^8$B neutrino flux via coherent elastic neutrino-nucleus scattering (CE$\nu$NS), and the $pp$ neutrino flux via neutrino-electron elastic scattering. 
With 200\,tonne$\cdot$year exposure, the number of detected  $^8$B neutrino CE$\nu$NS events is approximately 710, conservatively estimated based on the PandaX-4T efficiency curve in a specialized low energy analysis in Ref.~\cite{PandaX:2022aac}, with an energy threshold of $\sim$0.95\,keV$_{\textrm{nr}}$ (1\% efficiency) and the energy smearing taken into account. 
With an additional assumption of an equal number of accidental background events, PandaX-xT is capable of making a better-than 10\% precision measurement on the $^{8}$B neutrino fluxes (Fig.~\ref{fig:b8_flux}). We expect the efficiency and accidental background to be further improved along the way. 
For the $pp$ neutrinos, assuming a 2\% systematic uncertainty for material, krypton and radon background, the relative precision of $pp$ flux can be determined to 3\% using ER spectra fitting, 
better than the current best measurement from Borexino~\cite{Kumaran:2021lvv}.

\begin{figure*}[hbt]
    \centering
    \includegraphics[width=0.45\textwidth]{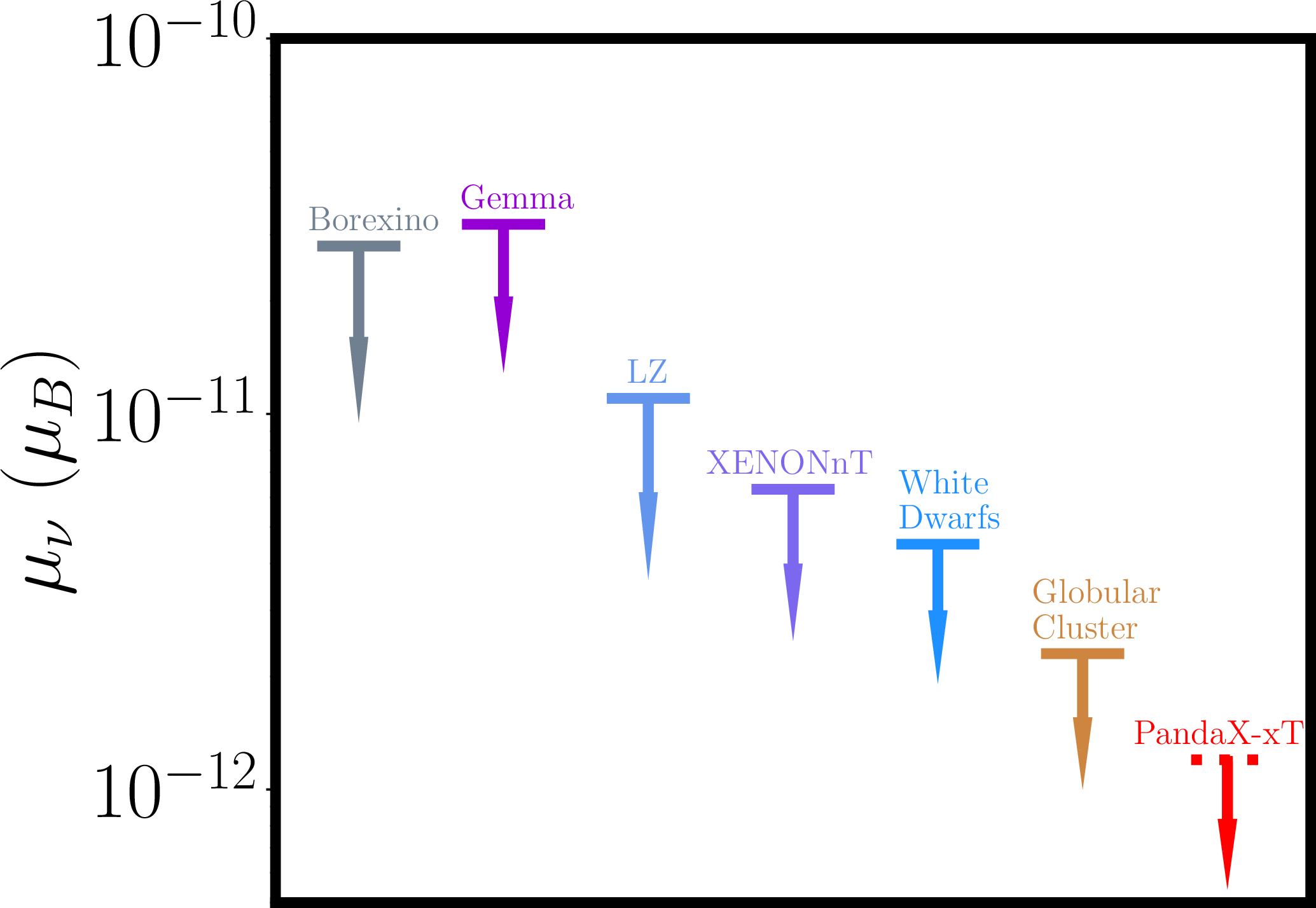}
    \caption{The sensitivity on the neutrino magnetic moment (90\% C.L) of PandaX-xT with 200\,tonne$\cdot$year exposure (red), together with  constraints from current leading experiments~\cite{Borexinodmumag, XENON:2022ltv, Gemmamumag,LZ:2023poo} and astrophysical observations~\cite{WhiteDwarfmumag,GlobularClustermumag}.
    }
    \label{fig:nv_sensitivity}
\end{figure*}

Note that the neutrino scattering rate in Tables~\ref{tab:bkg_below_10keV} and~\ref{tab:solar_nu_background} has assumed standard electroweak interaction only. If there is an enhanced neutrino magnetic moment $\mu_\nu$ due to the Majorana nature or exotic new physics beyond the SM, the elastic scattering cross section will be modified by a term~\cite{Billard:2013qya}
\begin{equation}
\left(\frac{d\sigma}{dT}\right)_{\mu_\nu} = \frac{\pi\alpha_{\rm EM}^2\mu_\nu^2}{m_e^2}\left[\frac{1-T/E_{\nu}}{T}\right]\,,
\end{equation}
in which $\alpha_{\rm EM} = 1/137$, $m_{e}$ is the mass of the electron, and $E_{\nu}$ and $T$ are the energy of the neutrino and recoiled electron, respectively. Therefore, the low energy $pp$ neutrino measurement can be in turn used to search for $\mu_\nu$. 
With 200\,tonne$\cdot$year exposure and our background presented in the ROI of DM in Table~\ref{tab:bkg_below_10keV}, we expect to achieve an unprecedented sensitivity of $1.2\times10^{-12}$$\mu_B$ to $\mu_\nu$ (Fig.~\ref{fig:nv_sensitivity}), where $\mu_B$ is the Bohr magneton.

\section{Discussions and Conclusions}\label{sec:conclusion}
This paper is a combination of letter-of-intent of PandaX-xT, a concise description of major design concepts, and physics potentials if the assumed performances are met. One should note that some of the technical choices presented still require significant R\&D. In Table~\ref{tab:matureness}, we summarize the readiness of individual subsystems and key techniques.

\begin{table*}[hbt]
    \centering
    {\scriptsize
    \begin{threeparttable}
    \caption{
    Project requirements and challenges. The \(\star\), \(\star\star\), and \(\star\star\star\) represent existing technology, what requires some R\&D but relies on well-established principles, and what is R\&D with uncertain outcome, respectively. 
    }
    \label{tab:matureness}
    \begin{tabular}{@{}llllp{2mm}}
        \toprule
        Item & Design goal & Current status & Challenge level \\ 
        \midrule
        Low background cryostat & thin-walled Cu & Low background SS & \(\star\star\star\) \\
        LXe TPC & 2.6 m (\diameter) \(\times\) 3.0 m (H) & 1.2 m (\diameter) \(\times\) 1.2 m (H) & \(\star\star\) \\
        Photosensor array & 10 m$^2$, high QE, low noises, fast & 3 m$^2$, R11410 3" PMT & \(\star\) \\
        Readout electronics & Cold and low background & Custom digitizer, outside, 500 MS/s & \(\star\star\star\) \\
        Energy resolution at 2.5 MeV & 1\% & 1.9\% & \(\star\) \\
        $^{85}$Kr background & 2 nBq/kg &  60 nBq/kg & \(\star\star\) \\
        $^{222}$Rn background & 0.5 $\mu$Bq/kg & 5 $\mu$Bq/kg & \(\star\star\) \\
        Cryogenics \& recirculation & 500 slpm gas (2 lpm liquid) & 150 slpm gas & \(\star\star\) \\
        IVETO & 200 keV threshold & under R\&D & \(\star\star\star\) \\ \bottomrule
    \end{tabular}
    \end{threeparttable}
    }
\end{table*}

Besides the standard methods of background rejection through data selection cuts, new methods are also being pursued by the PandaX collaboration. We anticipate improved measurement of the scintillation timing profiles for different recoil events~\cite{PhysRevB.27.5279, LUX:2018zdm, Hogenbirk:2018zwf} by utilizing the fast time response of the R12699.
In the spatial dimension, we plan to reconstruct individual scattering vertices to a sub-centimeter resolution, and track features in $\beta$s will be explored, to reduce the background in the MeV energy region. Taking further advantage of the photosensors with $\sim$ns timing response, we plan to investigate the detection of Cherenkov photons using both temporal and spatial information, a new method for background discrimination in the MeV region~\cite{Brodsky:2018abk}.

CJPL is being transformed into a mature national facility~\cite{Li:2014rca}, and most of the infrastructure from PandaX-4T will be inherited. These will ease the deployment of PandaX-xT.
The CJPL site provides additional key advantages to PandaX-xT. The large overburden gives a much lower cosmogenic background in comparison to laboratories at shallower sites, in particular $^{137}$Xe. The low latitude (28$^\circ$N) of CJPL and the higher field strength parallel to the earth's surface thereby suppresses low energy cosmic rays entering the atmosphere. This leads to a more than two times lower atmospheric neutrino background, the ``ultimate'' background for DM direct detection in the high mass region ($>$10~GeV/$c^2$), than those of high-latitude underground laboratories (LNGS, SURF, and SNOLAB)~\cite{Honda:2015fha,Zhuang:2021rsg}. Note that the prediction of the atmospheric neutrino flux and spectrum has rather large uncertainty. 
The PandaX-xT OVETO
could enable a measurement of atmospheric neutrino flux at CJPL.

PandaX-xT is envisioned to be carried out in multiple phases. In the first phase, natural xenon will be used as the detection target. We will gradually upgrade the detector based on the xenon in possession, reach a total target mass of 43~tonnes and operate for more than 10 years.
In the next phase, we envision that isotopic separation technology will be used to produce xenon with artificially modified isotopic abundance (AMIA), either via a split of odd and even nuclei~\cite{Suzuki:2000ch}, or further enrichment of $^{136}$Xe. 
Two functionally identical detectors can be filled with different AMIA xenon and operated simultaneously.
Two identical detectors but with different nuclear spins allows one to explore the SI and SD nature of the DM interactions.
The sensitivity to NLDBD can also be greatly improved by this ``dual-detector mode''. 

To summarize, we have demonstrated that the proposed PandaX-xT detector, with a 43~tonne active xenon target, will be a cutting-edge and multi-purpose observatory covering a wide range of physics topics in particle physics and astrophysics. With an exposure of 200\,tonne$\cdot$year, it allows a search sensitivity for DM-nucleon interactions to the neutrino floor~\cite{Billard:2013qya,OHare:2021utq,Schumann:2015cpa}, a key test on the WIMP DM. The same detector, operated for 10 years, will provide stringent tests on the Majorana nature of neutrinos using the NLDBD of $^{136}$Xe, effectively covering most of the parameter space for inverted neutrino MO. With its uniqueness in detecting low energy recoils, PandaX-xT will also allow precise detection of solar $^8$B and $pp$ neutrinos, as well as other neutrinos with astrophysical origins. These neutrinos can be used to broadly search for new physics, as well as in the context of  multi-messenger astronomy, such as detecting supernova explosions.

\section{Acknowledgement}
This project is supported in part by the Office of Science and Technology, Shanghai Municipal Government (grant Nos. 23JC1410200 and 22JCJC1410200), the Ministry of Science and Technology of China (grant No. 2023YFA1606200), and the National Science Foundation of China (grant Nos. 1209060, 12005131, 11905128, 11925502, 12222505, 12175139). 
We thank support from Double First Class Plan of the Shanghai Jiao Tong University, and the Tsung-Dao Lee Institute Experimental Platform Development Fund. We also thank the sponsorship from the Hongwen Foundation in Hong Kong, Tencent and New Cornerstone Science Foundation in China, and Yangyang Development Fund. Finally, we thank the CJPL administration and the Yalong River Hydropower Development Company Ltd. for indispensable logistical support and other help.

\newpage
\bibliographystyle{elsarticle-num}
\bibliography{references.bib}

\begin{thebibliography}{100}
\expandafter\ifx\csname url\endcsname\relax
  \def\url#1{\texttt{#1}}\fi
\expandafter\ifx\csname urlprefix\endcsname\relax\def\urlprefix{URL }\fi
\expandafter\ifx\csname href\endcsname\relax
  \def\href#1#2{#2} \def\path#1{#1}\fi

\bibitem{Bertone:2004pz}
G.~Bertone, D.~Hooper, J.~Silk, {Particle dark matter: Evidence, candidates and
  constraints}, Phys. Rept. 405 (2005) 279--390.
\newblock \href {http://arxiv.org/abs/hep-ph/0404175}
  {\path{arXiv:hep-ph/0404175}}, \href
  {https://doi.org/10.1016/j.physrep.2004.08.031}
  {\path{doi:10.1016/j.physrep.2004.08.031}}.

\bibitem{Jungman:1995df}
G.~Jungman, M.~Kamionkowski, K.~Griest, {Supersymmetric dark matter}, Phys.
  Rept. 267 (1996) 195--373.
\newblock \href {http://arxiv.org/abs/hep-ph/9506380}
  {\path{arXiv:hep-ph/9506380}}, \href
  {https://doi.org/10.1016/0370-1573(95)00058-5}
  {\path{doi:10.1016/0370-1573(95)00058-5}}.

\bibitem{Goodman:1984dc}
M.~W. Goodman, E.~Witten, {Detectability of Certain Dark Matter Candidates},
  Phys. Rev. D 31 (1985) 3059.
\newblock \href {https://doi.org/10.1103/PhysRevD.31.3059}
  {\path{doi:10.1103/PhysRevD.31.3059}}.

\bibitem{Gaitskell:2004gd}
R.~J. Gaitskell, {Direct detection of dark matter}, Ann. Rev. Nucl. Part. Sci.
  54 (2004) 315--359.
\newblock \href {https://doi.org/10.1146/annurev.nucl.54.070103.181244}
  {\path{doi:10.1146/annurev.nucl.54.070103.181244}}.

\bibitem{MarrodanUndagoitia:2015veg}
T.~Marrod\'an~Undagoitia, L.~Rauch, {Dark matter direct-detection experiments},
  J. Phys. G 43~(1) (2016) 013001.
\newblock \href {http://arxiv.org/abs/1509.08767} {\path{arXiv:1509.08767}},
  \href {https://doi.org/10.1088/0954-3899/43/1/013001}
  {\path{doi:10.1088/0954-3899/43/1/013001}}.

\bibitem{Liu:2017drf}
J.~Liu, X.~Chen, X.~Ji, {Current status of direct dark matter detection
  experiments}, Nature Phys. 13~(3) (2017) 212--216.
\newblock \href {http://arxiv.org/abs/1709.00688} {\path{arXiv:1709.00688}},
  \href {https://doi.org/10.1038/nphys4039} {\path{doi:10.1038/nphys4039}}.

\bibitem{Schumann:2019eaa}
M.~Schumann, {Direct Detection of WIMP Dark Matter: Concepts and Status}, J.
  Phys. G 46~(10) (2019) 103003.
\newblock \href {http://arxiv.org/abs/1903.03026} {\path{arXiv:1903.03026}},
  \href {https://doi.org/10.1088/1361-6471/ab2ea5}
  {\path{doi:10.1088/1361-6471/ab2ea5}}.

\bibitem{Billard:2021uyg}
J.~Billard, et~al., {Direct detection of dark matter\textemdash{}APPEC
  committee report}, Rept. Prog. Phys. 85~(5) (2022) 056201.
\newblock \href {http://arxiv.org/abs/2104.07634} {\path{arXiv:2104.07634}},
  \href {https://doi.org/10.1088/1361-6633/ac5754}
  {\path{doi:10.1088/1361-6633/ac5754}}.

\bibitem{Cushman:2013zza}
P.~Cushman, et~al., {Working Group Report: WIMP Dark Matter Direct Detection},
  in: {Snowmass 2013}: {Snowmass on the Mississippi}, 2013.
\newblock \href {http://arxiv.org/abs/1310.8327} {\path{arXiv:1310.8327}}.

\bibitem{PandaX-4T:2021bab}
Y.~Meng, et~al., {Dark Matter Search Results from the PandaX-4T Commissioning
  Run}, Phys. Rev. Lett. 127~(26) (2021) 261802.
\newblock \href {http://arxiv.org/abs/2107.13438} {\path{arXiv:2107.13438}},
  \href {https://doi.org/10.1103/PhysRevLett.127.261802}
  {\path{doi:10.1103/PhysRevLett.127.261802}}.

\bibitem{XENON:2023cxc}
E.~Aprile, et~al., {First Dark Matter Search with Nuclear Recoils from the
  XENONnT Experiment}, Phys. Rev. Lett. 131~(4) (2023) 041003.
\newblock \href {http://arxiv.org/abs/2303.14729} {\path{arXiv:2303.14729}},
  \href {https://doi.org/10.1103/PhysRevLett.131.041003}
  {\path{doi:10.1103/PhysRevLett.131.041003}}.

\bibitem{LZ:2022lsv}
J.~Aalbers, et~al., {First Dark Matter Search Results from the LUX-ZEPLIN (LZ)
  Experiment}, Phys. Rev. Lett. 131~(4) (2023) 041002.
\newblock \href {http://arxiv.org/abs/2207.03764} {\path{arXiv:2207.03764}},
  \href {https://doi.org/10.1103/PhysRevLett.131.041002}
  {\path{doi:10.1103/PhysRevLett.131.041002}}.

\bibitem{Billard:2013qya}
J.~Billard, L.~Strigari, E.~Figueroa-Feliciano, {Implication of neutrino
  backgrounds on the reach of next generation dark matter direct detection
  experiments}, Phys. Rev. D 89~(2) (2014) 023524.
\newblock \href {http://arxiv.org/abs/1307.5458} {\path{arXiv:1307.5458}},
  \href {https://doi.org/10.1103/PhysRevD.89.023524}
  {\path{doi:10.1103/PhysRevD.89.023524}}.

\bibitem{OHare:2021utq}
C.~A.~J. O'Hare, {New Definition of the Neutrino Floor for Direct Dark Matter
  Searches}, Phys. Rev. Lett. 127~(25) (2021) 251802.
\newblock \href {http://arxiv.org/abs/2109.03116} {\path{arXiv:2109.03116}},
  \href {https://doi.org/10.1103/PhysRevLett.127.251802}
  {\path{doi:10.1103/PhysRevLett.127.251802}}.

\bibitem{Majorana:1937vz}
E.~Majorana, {Teoria simmetrica dell\textquoteright{}elettrone e del
  positrone}, Nuovo Cim. 14 (1937) 171--184.
\newblock \href {https://doi.org/10.1007/BF02961314}
  {\path{doi:10.1007/BF02961314}}.

\bibitem{Minkowski:1977sc}
P.~Minkowski, {$\mu \to e\gamma$ at a Rate of One Out of $10^{9}$ Muon
  Decays?}, Phys. Lett. B 67 (1977) 421--428.
\newblock \href {https://doi.org/10.1016/0370-2693(77)90435-X}
  {\path{doi:10.1016/0370-2693(77)90435-X}}.

\bibitem{yanagida1979proc}
T.~Yanagida, Proc. workshop on unified theory and the baryon number in the
  universe, KEK Report No. 79-18 95 (1979).

\bibitem{GellMann:1980vs}
M.~Gell-Mann, P.~Ramond, R.~Slansky, {Complex Spinors and Unified Theories},
  Conf. Proc. C 790927 (1979) 315--321.
\newblock \href {http://arxiv.org/abs/1306.4669} {\path{arXiv:1306.4669}}.

\bibitem{Glashow:1979nm}
S.~L. Glashow, {The Future of Elementary Particle Physics}, NATO Sci. Ser. B 61
  (1980) 687.
\newblock \href {https://doi.org/10.1007/978-1-4684-7197-7_15}
  {\path{doi:10.1007/978-1-4684-7197-7_15}}.

\bibitem{Mohapatra:1979ia}
R.~N. Mohapatra, G.~Senjanovic, {Neutrino Mass and Spontaneous Parity
  Nonconservation}, Phys. Rev. Lett. 44 (1980) 912.
\newblock \href {https://doi.org/10.1103/PhysRevLett.44.912}
  {\path{doi:10.1103/PhysRevLett.44.912}}.

\bibitem{Fukugita:1986hr}
M.~Fukugita, T.~Yanagida, {Baryogenesis Without Grand Unification}, Phys. Lett.
  B 174 (1986) 45--47.
\newblock \href {https://doi.org/10.1016/0370-2693(86)91126-3}
  {\path{doi:10.1016/0370-2693(86)91126-3}}.

\bibitem{Furry:1939qr}
W.~H. Furry, {On transition probabilities in double beta-disintegration}, Phys.
  Rev. 56 (1939) 1184--1193.
\newblock \href {https://doi.org/10.1103/PhysRev.56.1184}
  {\path{doi:10.1103/PhysRev.56.1184}}.

\bibitem{Agostini:2022zub}
M.~Agostini, G.~Benato, J.~A. Detwiler, J.~Men\'endez, F.~Vissani, {Toward the
  discovery of matter creation with neutrinoless
  \ensuremath{\beta}\ensuremath{\beta} decay}, Rev. Mod. Phys. 95~(2) (2023)
  025002.
\newblock \href {http://arxiv.org/abs/2202.01787} {\path{arXiv:2202.01787}},
  \href {https://doi.org/10.1103/RevModPhys.95.025002}
  {\path{doi:10.1103/RevModPhys.95.025002}}.

\bibitem{GERDA:2020xhi}
M.~Agostini, et~al., {Final Results of GERDA on the Search for Neutrinoless
  Double-$\beta$ Decay}, Phys. Rev. Lett. 125~(25) (2020) 252502.
\newblock \href {http://arxiv.org/abs/2009.06079} {\path{arXiv:2009.06079}},
  \href {https://doi.org/10.1103/PhysRevLett.125.252502}
  {\path{doi:10.1103/PhysRevLett.125.252502}}.

\bibitem{Majorana:2022udl}
I.~J. Arnquist, et~al., {Final Result of the Majorana
  Demonstrator\textquoteright{}s Search for Neutrinoless
  Double-\ensuremath{\beta} Decay in Ge76}, Phys. Rev. Lett. 130~(6) (2023)
  062501.
\newblock \href {http://arxiv.org/abs/2207.07638} {\path{arXiv:2207.07638}},
  \href {https://doi.org/10.1103/PhysRevLett.130.062501}
  {\path{doi:10.1103/PhysRevLett.130.062501}}.

\bibitem{EXO-200:2019rkq}
G.~Anton, et~al., {Search for Neutrinoless Double-$\beta$ Decay with the
  Complete EXO-200 Dataset}, Phys. Rev. Lett. 123~(16) (2019) 161802.
\newblock \href {http://arxiv.org/abs/1906.02723} {\path{arXiv:1906.02723}},
  \href {https://doi.org/10.1103/PhysRevLett.123.161802}
  {\path{doi:10.1103/PhysRevLett.123.161802}}.

\bibitem{CUORE:2021mvw}
D.~Q. Adams, et~al., {Search for Majorana neutrinos exploiting millikelvin
  cryogenics with CUORE}, Nature 604~(7904) (2022) 53--58.
\newblock \href {http://arxiv.org/abs/2104.06906} {\path{arXiv:2104.06906}},
  \href {https://doi.org/10.1038/s41586-022-04497-4}
  {\path{doi:10.1038/s41586-022-04497-4}}.

\bibitem{SNO:2021xpa}
V.~Albanese, et~al., {The SNO+ experiment}, JINST 16~(08) (2021) P08059.
\newblock \href {http://arxiv.org/abs/2104.11687} {\path{arXiv:2104.11687}},
  \href {https://doi.org/10.1088/1748-0221/16/08/P08059}
  {\path{doi:10.1088/1748-0221/16/08/P08059}}.

\bibitem{NEMO-3:2015jgm}
R.~Arnold, et~al., {Results of the search for neutrinoless
  double-\ensuremath{\beta} decay in $^{100}$Mo with the NEMO-3 experiment},
  Phys. Rev. D 92~(7) (2015) 072011.
\newblock \href {http://arxiv.org/abs/1506.05825} {\path{arXiv:1506.05825}},
  \href {https://doi.org/10.1103/PhysRevD.92.072011}
  {\path{doi:10.1103/PhysRevD.92.072011}}.

\bibitem{KamLAND-Zen:2022tow}
S.~Abe, et~al., {Search for the Majorana Nature of Neutrinos in the Inverted
  Mass Ordering Region with KamLAND-Zen}, Phys. Rev. Lett. 130~(5) (2023)
  051801.
\newblock \href {http://arxiv.org/abs/2203.02139} {\path{arXiv:2203.02139}},
  \href {https://doi.org/10.1103/PhysRevLett.130.051801}
  {\path{doi:10.1103/PhysRevLett.130.051801}}.

\bibitem{nEXO:2021ujk}
G.~Adhikari, et~al., {nEXO: neutrinoless double beta decay search beyond
  10$^{28}$ year half-life sensitivity}, J. Phys. G 49~(1) (2022) 015104.
\newblock \href {http://arxiv.org/abs/2106.16243} {\path{arXiv:2106.16243}},
  \href {https://doi.org/10.1088/1361-6471/ac3631}
  {\path{doi:10.1088/1361-6471/ac3631}}.

\bibitem{Nakamura:2020szx}
R.~Nakamura, H.~Sambonsugi, K.~Shiraishi, Y.~Wada, {Research and development
  toward KamLAND2-Zen}, J. Phys. Conf. Ser. 1468~(1) (2020) 012256.
\newblock \href {https://doi.org/10.1088/1742-6596/1468/1/012256}
  {\path{doi:10.1088/1742-6596/1468/1/012256}}.

\bibitem{Brugnera:2023zgw}
R.~Brugnera, {Neutrinoless double beta decay search with LEGEND}, PoS NOW2022
  (2023) 075.
\newblock \href {https://doi.org/10.22323/1.421.0075}
  {\path{doi:10.22323/1.421.0075}}.

\bibitem{Ma:2023yrk}
H.~Ma, W.~Dai, L.~Yang, {CDEX-300\ensuremath{\nu} program for Ge-76
  neutrinoless double beta decay search}, PoS TAUP2023 (2024) 200.
\newblock \href {https://doi.org/10.22323/1.441.0200}
  {\path{doi:10.22323/1.441.0200}}.

\bibitem{Zhao:2016brs}
J.~Zhao, L.-J. Wen, Y.-F. Wang, J.~Cao, {Physics potential of searching for
  $0\nu\beta\beta$ decays in JUNO}, Chin. Phys. C 41~(5) (2017) 053001.
\newblock \href {http://arxiv.org/abs/1610.07143} {\path{arXiv:1610.07143}},
  \href {https://doi.org/10.1088/1674-1137/41/5/053001}
  {\path{doi:10.1088/1674-1137/41/5/053001}}.

\bibitem{CUPID:2022jlk}
K.~Alfonso, et~al., {CUPID: The Next-Generation Neutrinoless Double Beta Decay
  Experiment}, J. Low Temp. Phys. 211~(5-6) (2023) 375--383.
\newblock \href {https://doi.org/10.1007/s10909-022-02909-3}
  {\path{doi:10.1007/s10909-022-02909-3}}.

\bibitem{XENON:2020iwh}
E.~Aprile, et~al., {Energy resolution and linearity of XENON1T in the MeV
  energy range}, Eur. Phys. J. C 80~(8) (2020) 785.
\newblock \href {http://arxiv.org/abs/2003.03825} {\path{arXiv:2003.03825}},
  \href {https://doi.org/10.1140/epjc/s10052-020-8284-0}
  {\path{doi:10.1140/epjc/s10052-020-8284-0}}.

\bibitem{Pereira:2023rte}
G.~Pereira, C.~Silva, V.~N. Solovov, {Energy resolution of the LZ detector for
  high-energy electronic recoils}, JINST 18~(04) (2023) C04007.
\newblock \href {https://doi.org/10.1088/1748-0221/18/04/C04007}
  {\path{doi:10.1088/1748-0221/18/04/C04007}}.

\bibitem{Aalbers:2022dzr}
J.~Aalbers, et~al., {A next-generation liquid xenon observatory for dark matter
  and neutrino physics}, J. Phys. G 50~(1) (2023) 013001.
\newblock \href {http://arxiv.org/abs/2203.02309} {\path{arXiv:2203.02309}},
  \href {https://doi.org/10.1088/1361-6471/ac841a}
  {\path{doi:10.1088/1361-6471/ac841a}}.

\bibitem{DARWIN:2016hyl}
J.~Aalbers, et~al., {DARWIN: towards the ultimate dark matter detector}, JCAP
  11 (2016) 017.
\newblock \href {http://arxiv.org/abs/1606.07001} {\path{arXiv:1606.07001}},
  \href {https://doi.org/10.1088/1475-7516/2016/11/017}
  {\path{doi:10.1088/1475-7516/2016/11/017}}.

\bibitem{XLZD}
\href{https://darwin.physik.uzh.ch/news.html, https://xlzd.org}{{XLZD: Joining
  forces towards a next-generation Dark Matter experiment}},
  \url{https://darwin.physik.uzh.ch/news.html, https://xlzd.org}.
\newline\urlprefix\url{https://darwin.physik.uzh.ch/news.html,
  https://xlzd.org}

\bibitem{Cheng:2017usi}
J.-P. Cheng, et~al., {The China Jinping Underground Laboratory and its Early
  Science}, Ann. Rev. Nucl. Part. Sci. 67 (2017) 231--251.
\newblock \href {http://arxiv.org/abs/1801.00587} {\path{arXiv:1801.00587}},
  \href {https://doi.org/10.1146/annurev-nucl-102115-044842}
  {\path{doi:10.1146/annurev-nucl-102115-044842}}.

\bibitem{PandaX:2014mem}
X.~Cao, et~al., {PandaX: A Liquid Xenon Dark Matter Experiment at CJPL}, Sci.
  China Phys. Mech. Astron. 57 (2014) 1476--1494.
\newblock \href {http://arxiv.org/abs/1405.2882} {\path{arXiv:1405.2882}},
  \href {https://doi.org/10.1007/s11433-014-5521-2}
  {\path{doi:10.1007/s11433-014-5521-2}}.

\bibitem{PandaX:2018wtu}
H.~Zhang, et~al., {Dark matter direct search sensitivity of the PandaX-4T
  experiment}, Sci. China Phys. Mech. Astron. 62~(3) (2019) 31011.
\newblock \href {http://arxiv.org/abs/1806.02229} {\path{arXiv:1806.02229}},
  \href {https://doi.org/10.1007/s11433-018-9259-0}
  {\path{doi:10.1007/s11433-018-9259-0}}.

\bibitem{CDEX:2013kpt}
K.-J. Kang, et~al., {Introduction to the CDEX experiment}, Front. Phys.
  (Beijing) 8 (2013) 412--437.
\newblock \href {http://arxiv.org/abs/1303.0601} {\path{arXiv:1303.0601}},
  \href {https://doi.org/10.1007/s11467-013-0349-1}
  {\path{doi:10.1007/s11467-013-0349-1}}.

\bibitem{CDEX:2014amu}
Q.~Yue, et~al., {Limits on light WIMPs from the CDEX-1 experiment with a p-type
  point-contact germanium detector at the China Jingping Underground
  Laboratory}, Phys. Rev. D 90 (2014) 091701.
\newblock \href {http://arxiv.org/abs/1404.4946} {\path{arXiv:1404.4946}},
  \href {https://doi.org/10.1103/PhysRevD.90.091701}
  {\path{doi:10.1103/PhysRevD.90.091701}}.

\bibitem{CDEX:2018lau}
H.~Jiang, et~al., {Limits on Light Weakly Interacting Massive Particles from
  the First 102.8 kg ${\times}$ day Data of the CDEX-10 Experiment}, Phys. Rev.
  Lett. 120~(24) (2018) 241301.
\newblock \href {http://arxiv.org/abs/1802.09016} {\path{arXiv:1802.09016}},
  \href {https://doi.org/10.1103/PhysRevLett.120.241301}
  {\path{doi:10.1103/PhysRevLett.120.241301}}.

\bibitem{PandaX:2014ria}
M.~Xiao, et~al., {First dark matter search results from the PandaX-I
  experiment}, Sci. China Phys. Mech. Astron. 57 (2014) 2024--2030.
\newblock \href {http://arxiv.org/abs/1408.5114} {\path{arXiv:1408.5114}},
  \href {https://doi.org/10.1007/s11433-014-5598-7}
  {\path{doi:10.1007/s11433-014-5598-7}}.

\bibitem{PandaX-II:2016andi}
A.~Tan, et~al., {Dark Matter Results from First 98.7 Days of Data from the
  PandaX-II Experiment}, Phys. Rev. Lett. 117~(12) (2016) 121303.
\newblock \href {http://arxiv.org/abs/1607.07400} {\path{arXiv:1607.07400}},
  \href {https://doi.org/10.1103/PhysRevLett.117.121303}
  {\path{doi:10.1103/PhysRevLett.117.121303}}.

\bibitem{PandaX-II:2017hlx}
X.~Cui, et~al., {Dark Matter Results From 54-Ton-Day Exposure of PandaX-II
  Experiment}, Phys. Rev. Lett. 119~(18) (2017) 181302.
\newblock \href {http://arxiv.org/abs/1708.06917} {\path{arXiv:1708.06917}},
  \href {https://doi.org/10.1103/PhysRevLett.119.181302}
  {\path{doi:10.1103/PhysRevLett.119.181302}}.

\bibitem{PandaX-II:2020udv}
X.~Zhou, et~al., {A Search for Solar Axions and Anomalous Neutrino Magnetic
  Moment with the Complete PandaX-II Data}, Chin. Phys. Lett. 38~(1) (2021)
  011301.
\newblock \href {http://arxiv.org/abs/2008.06485} {\path{arXiv:2008.06485}},
  \href {https://doi.org/10.1088/0256-307X/38/10/109902}
  {\path{doi:10.1088/0256-307X/38/10/109902}}.

\bibitem{Li:2014rca}
J.~Li, X.~Ji, W.~Haxton, J.~S.~Y. Wang, {The second-phase development of the
  China JinPing underground Laboratory}, Phys. Procedia 61 (2015) 576--585.
\newblock \href {https://doi.org/10.1016/j.phpro.2014.12.055}
  {\path{doi:10.1016/j.phpro.2014.12.055}}.

\bibitem{PandaX:2022xas}
Z.~Huang, et~al., {Constraints on the axial-vector and pseudo-scalar mediated
  WIMP-nucleus interactions from PandaX-4T experiment}, Phys. Lett. B 834
  (2022) 137487.
\newblock \href {http://arxiv.org/abs/2208.03626} {\path{arXiv:2208.03626}},
  \href {https://doi.org/10.1016/j.physletb.2022.137487}
  {\path{doi:10.1016/j.physletb.2022.137487}}.

\bibitem{PandaX:2022xqx}
S.~Li, et~al., {Search for Light Dark Matter with Ionization Signals in the
  PandaX-4T Experiment}, Phys. Rev. Lett. 130~(26) (2023) 261001.
\newblock \href {http://arxiv.org/abs/2212.10067} {\path{arXiv:2212.10067}},
  \href {https://doi.org/10.1103/PhysRevLett.130.261001}
  {\path{doi:10.1103/PhysRevLett.130.261001}}.

\bibitem{PandaX:2023xgl}
D.~Huang, et~al., {Search for Dark-Matter\textendash{}Nucleon Interactions with
  a Dark Mediator in PandaX-4T}, Phys. Rev. Lett. 131~(19) (2023) 191002.
\newblock \href {http://arxiv.org/abs/2308.01540} {\path{arXiv:2308.01540}},
  \href {https://doi.org/10.1103/PhysRevLett.131.191002}
  {\path{doi:10.1103/PhysRevLett.131.191002}}.

\bibitem{PandaX:2023toi}
X.~Ning, et~al., {Limits on the luminance of dark matter from xenon recoil
  data}, Nature 618~(7963) (2023) 47--50.
\newblock \href {https://doi.org/10.1038/s41586-023-05982-0}
  {\path{doi:10.1038/s41586-023-05982-0}}.

\bibitem{BaiYangComment}
Y.~Bai, {Dark Matter is Darker}, NUCLEAR SCIENCE AND TECHNIQUES 34~(6) (2023)
  76.
\newblock \href {https://doi.org/10.1007/s41365-023-01249-5}
  {\path{doi:10.1007/s41365-023-01249-5}}.

\bibitem{PandaX:2022kwg}
L.~Si, et~al., {Determination of Double Beta Decay Half-Life of 136Xe with the
  PandaX-4T Natural Xenon Detector}, Research 2022 (2022) 9798721.
\newblock \href {http://arxiv.org/abs/2205.12809} {\path{arXiv:2205.12809}},
  \href {https://doi.org/10.34133/2022/9798721}
  {\path{doi:10.34133/2022/9798721}}.

\bibitem{PandaX:2023ggs}
X.~Yan, et~al., {Searching for Two-Neutrino and Neutrinoless Double Beta Decay
  of Xe134 with the PandaX-4T Experiment}, Phys. Rev. Lett. 132~(15) (2024)
  152502.
\newblock \href {http://arxiv.org/abs/2312.15632} {\path{arXiv:2312.15632}},
  \href {https://doi.org/10.1103/PhysRevLett.132.152502}
  {\path{doi:10.1103/PhysRevLett.132.152502}}.

\bibitem{PandaX-4T:2021lbm}
Z.~Qian, et~al., {Low radioactive material screening and background control for
  the PandaX-4T experiment}, JHEP 06 (2022) 147.
\newblock \href {http://arxiv.org/abs/2112.02892} {\path{arXiv:2112.02892}},
  \href {https://doi.org/10.1007/JHEP06(2022)147}
  {\path{doi:10.1007/JHEP06(2022)147}}.

\bibitem{Ma_2021}
H.~Ma, W.~Dai, Z.~Zeng, T.~Xue, L.~Yang, Q.~Yue, J.~Cheng,
  \href{https://dx.doi.org/10.1088/1742-6596/2156/1/012170}{Status and prospect
  of china jinping underground laboratory}, Journal of Physics: Conference
  Series 2156~(1) (2021) 012170.
\newblock \href {https://doi.org/10.1088/1742-6596/2156/1/012170}
  {\path{doi:10.1088/1742-6596/2156/1/012170}}.
\newline\urlprefix\url{https://dx.doi.org/10.1088/1742-6596/2156/1/012170}

\bibitem{GEANT4:2002zbu}
S.~Agostinelli, et~al., {GEANT4--a simulation toolkit}, Nucl. Instrum. Meth. A
  506 (2003) 250--303.
\newblock \href {https://doi.org/10.1016/S0168-9002(03)01368-8}
  {\path{doi:10.1016/S0168-9002(03)01368-8}}.

\bibitem{Yeh:2011zz}
M.~Yeh, S.~Hans, W.~Beriguete, R.~Rosero, L.~Hu, R.~L. Hahn, M.~V. Diwan, D.~E.
  Jaffe, S.~H. Kettell, L.~Littenberg, {A new water-based liquid scintillator
  and potential applications}, Nucl. Instrum. Meth. A 660 (2011) 51--56.
\newblock \href {https://doi.org/10.1016/j.nima.2011.08.040}
  {\path{doi:10.1016/j.nima.2011.08.040}}.

\bibitem{Alonso:2014fwf}
J.~R. Alonso, et~al., {Advanced Scintillator Detector Concept (ASDC): A Concept
  Paper on the Physics Potential of Water-Based Liquid Scintillator} (9 2014).
\newblock \href {http://arxiv.org/abs/1409.5864} {\path{arXiv:1409.5864}}.

\bibitem{Bignell:2015oqa}
L.~J. Bignell, et~al., {Characterization and Modeling of a Water-based Liquid
  Scintillator}, JINST 10~(12) (2015) P12009.
\newblock \href {http://arxiv.org/abs/1508.07029} {\path{arXiv:1508.07029}},
  \href {https://doi.org/10.1088/1748-0221/10/12/P12009}
  {\path{doi:10.1088/1748-0221/10/12/P12009}}.

\bibitem{Theia:2019non}
M.~Askins, et~al., {THEIA: an advanced optical neutrino detector}, Eur. Phys.
  J. C 80~(5) (2020) 416.
\newblock \href {http://arxiv.org/abs/1911.03501} {\path{arXiv:1911.03501}},
  \href {https://doi.org/10.1140/epjc/s10052-020-7977-8}
  {\path{doi:10.1140/epjc/s10052-020-7977-8}}.

\bibitem{Land:2020oiz}
B.~J. Land, Z.~Bagdasarian, J.~Caravaca, M.~Smiley, M.~Yeh, G.~D. Orebi~Gann,
  {MeV-scale performance of water-based and pure liquid scintillator
  detectors}, Phys. Rev. D 103~(5) (2021) 052004.
\newblock \href {http://arxiv.org/abs/2007.14999} {\path{arXiv:2007.14999}},
  \href {https://doi.org/10.1103/PhysRevD.103.052004}
  {\path{doi:10.1103/PhysRevD.103.052004}}.

\bibitem{Theia:2022uyh}
M.~Askins, et~al., {Theia: Summary of physics program. Snowmass White Paper
  Submission}, in: {2022 Snowmass Summer Study}, 2022.
\newblock \href {http://arxiv.org/abs/2202.12839} {\path{arXiv:2202.12839}}.

\bibitem{Suzuki:2019jby}
Y.~Suzuki, {The Super-Kamiokande experiment}, Eur. Phys. J. C 79~(4) (2019)
  298.
\newblock \href {https://doi.org/10.1140/epjc/s10052-019-6796-2}
  {\path{doi:10.1140/epjc/s10052-019-6796-2}}.

\bibitem{Zhuang:2021rsg}
Y.~Zhuang, L.~E. Strigari, R.~F. Lang, {Time variation of the atmospheric
  neutrino flux at dark matter detectors}, Phys. Rev. D 105~(4) (2022) 043001.
\newblock \href {http://arxiv.org/abs/2110.14723} {\path{arXiv:2110.14723}},
  \href {https://doi.org/10.1103/PhysRevD.105.043001}
  {\path{doi:10.1103/PhysRevD.105.043001}}.

\bibitem{JUNO:2020ijm}
A.~Abusleme, et~al., {TAO Conceptual Design Report: A Precision Measurement of
  the Reactor Antineutrino Spectrum with Sub-percent Energy Resolution} (5
  2020).
\newblock \href {http://arxiv.org/abs/2005.08745} {\path{arXiv:2005.08745}}.

\bibitem{Xie:2020bqa}
Z.~Xie, J.~Cao, Y.~Ding, M.~Liu, X.~Sun, W.~Wang, Y.~Xie, {A liquid
  scintillator for a neutrino detector working at \ensuremath{-}50 degree},
  Nucl. Instrum. Meth. A 1009 (2021) 165459.
\newblock \href {http://arxiv.org/abs/2012.11883} {\path{arXiv:2012.11883}},
  \href {https://doi.org/10.1016/j.nima.2021.165459}
  {\path{doi:10.1016/j.nima.2021.165459}}.

\bibitem{MINOS:2002xlc}
P.~Adamson, et~al., {The MINOS scintillator calorimeter system}, IEEE Trans.
  Nucl. Sci. 49 (2002) 861--863.
\newblock \href {https://doi.org/10.1109/TNS.2002.1039579}
  {\path{doi:10.1109/TNS.2002.1039579}}.

\bibitem{NOvA:2007rmc}
D.~S. Ayres, et~al., {The NOvA Technical Design Report} (10 2007).
\newblock \href {https://doi.org/10.2172/935497} {\path{doi:10.2172/935497}}.

\bibitem{GERDA:2017ihb}
M.~Agostini, et~al., {Upgrade for Phase II of the Gerda experiment}, Eur. Phys.
  J. C 78~(5) (2018) 388.
\newblock \href {http://arxiv.org/abs/1711.01452} {\path{arXiv:1711.01452}},
  \href {https://doi.org/10.1140/epjc/s10052-018-5812-2}
  {\path{doi:10.1140/epjc/s10052-018-5812-2}}.

\bibitem{Buck:2019tsa}
C.~Buck, B.~Gramlich, S.~Schoppmann, {Novel Opaque Scintillator for Neutrino
  Detection}, JINST 14~(11) (2019) P11007.
\newblock \href {http://arxiv.org/abs/1908.03334} {\path{arXiv:1908.03334}},
  \href {https://doi.org/10.1088/1748-0221/14/11/P11007}
  {\path{doi:10.1088/1748-0221/14/11/P11007}}.

\bibitem{Auger:2012gs}
M.~Auger, et~al., {The EXO-200 detector, part I: Detector design and
  construction}, JINST 7 (2012) P05010.
\newblock \href {http://arxiv.org/abs/1202.2192} {\path{arXiv:1202.2192}},
  \href {https://doi.org/10.1088/1748-0221/7/05/P05010}
  {\path{doi:10.1088/1748-0221/7/05/P05010}}.

\bibitem{Aprile:2014ila}
E.~Aprile, H.~Contreras, L.~W. Goetzke, A.~J. Melgarejo~Fernandez, M.~Messina,
  J.~Naganoma, G.~Plante, A.~Rizzo, P.~Shagin, R.~Wall, {Measurements of
  proportional scintillation and electron multiplication in liquid xenon using
  thin wires}, JINST 9~(11) (2014) P11012.
\newblock \href {http://arxiv.org/abs/1408.6206} {\path{arXiv:1408.6206}},
  \href {https://doi.org/10.1088/1748-0221/9/11/P11012}
  {\path{doi:10.1088/1748-0221/9/11/P11012}}.

\bibitem{Ye:2014gga}
T.~Ye, K.~L. Giboni, X.~Ji, {Initial evaluation of proportional scintillation
  in liquid Xenon for direct dark matter detection}, JINST 9~(12) (2014)
  P12007.
\newblock \href {https://doi.org/10.1088/1748-0221/9/12/P12007}
  {\path{doi:10.1088/1748-0221/9/12/P12007}}.

\bibitem{Juyal:2019gch}
P.~Juyal, K.-L. Giboni, X.-D. Ji, J.-L. Liu, {On proportional scintillation in
  very large liquid xenon detectors}, Nucl. Sci. Tech. 31~(9) (2020) 93.
\newblock \href {http://arxiv.org/abs/1910.13160} {\path{arXiv:1910.13160}},
  \href {https://doi.org/10.1007/s41365-020-00797-4}
  {\path{doi:10.1007/s41365-020-00797-4}}.

\bibitem{DEAP-3600:2017ker}
P.~A. Amaudruz, et~al., {Design and Construction of the DEAP-3600 Dark Matter
  Detector}, Astropart. Phys. 108 (2019) 1--23.
\newblock \href {http://arxiv.org/abs/1712.01982} {\path{arXiv:1712.01982}},
  \href {https://doi.org/10.1016/j.astropartphys.2018.09.006}
  {\path{doi:10.1016/j.astropartphys.2018.09.006}}.

\bibitem{JUNO:2021kxb}
A.~Abusleme, et~al., {Radioactivity control strategy for the JUNO detector},
  JHEP 11 (2021) 102.
\newblock \href {http://arxiv.org/abs/2107.03669} {\path{arXiv:2107.03669}},
  \href {https://doi.org/10.1007/JHEP11(2021)102}
  {\path{doi:10.1007/JHEP11(2021)102}}.

\bibitem{CAO2021165377}
C.~Cao, N.~Li, X.~Yang, J.~Zhao, Y.~Li, Z.~Cai, L.~Wen, X.~Luo, Y.~Heng,
  Y.~Ding,
  \href{https://www.sciencedirect.com/science/article/pii/S0168900221003612}{A
  practical approach of high precision u and th concentration measurement in
  acrylic}, Nuclear Instruments and Methods in Physics Research Section A:
  Accelerators, Spectrometers, Detectors and Associated Equipment 1004 (2021)
  165377.
\newblock \href {https://doi.org/https://doi.org/10.1016/j.nima.2021.165377}
  {\path{doi:https://doi.org/10.1016/j.nima.2021.165377}}.
\newline\urlprefix\url{https://www.sciencedirect.com/science/article/pii/S0168900221003612}

\bibitem{Althuser:2020uxc}
L.~Alth\"user, S.~Lindemann, M.~Murra, M.~Schumann, C.~Wittweg, C.~Weinheimer,
  {VUV Transmission of PTFE for Xenon-based Particle Detectors}, JINST 15~(12)
  (2020) P12021.
\newblock \href {http://arxiv.org/abs/2006.05827} {\path{arXiv:2006.05827}},
  \href {https://doi.org/10.1088/1748-0221/15/12/P12021}
  {\path{doi:10.1088/1748-0221/15/12/P12021}}.

\bibitem{Kastens:2009pa}
L.~W. Kastens, S.~B. Cahn, A.~Manzur, D.~N. McKinsey, {Calibration of a Liquid
  Xenon Detector with Kr-83m}, Phys. Rev. C 80 (2009) 045809.
\newblock \href {http://arxiv.org/abs/0905.1766} {\path{arXiv:0905.1766}},
  \href {https://doi.org/10.1103/PhysRevC.80.045809}
  {\path{doi:10.1103/PhysRevC.80.045809}}.

\bibitem{LUX:2015amk}
D.~S. Akerib, et~al., {Tritium calibration of the LUX dark matter experiment},
  Phys. Rev. D 93~(7) (2016) 072009.
\newblock \href {http://arxiv.org/abs/1512.03133} {\path{arXiv:1512.03133}},
  \href {https://doi.org/10.1103/PhysRevD.93.072009}
  {\path{doi:10.1103/PhysRevD.93.072009}}.

\bibitem{XENON:2016rze}
E.~Aprile, et~al., {Results from a Calibration of XENON100 Using a Source of
  Dissolved Radon-220}, Phys. Rev. D 95~(7) (2017) 072008.
\newblock \href {http://arxiv.org/abs/1611.03585} {\path{arXiv:1611.03585}},
  \href {https://doi.org/10.1103/PhysRevD.95.072008}
  {\path{doi:10.1103/PhysRevD.95.072008}}.

\bibitem{Ma:2020kll}
W.~Ma, et~al., {Internal calibration of the PandaX-II detector with radon
  gaseous sources}, JINST 15~(12) (2020) P12038.
\newblock \href {http://arxiv.org/abs/2006.09311} {\path{arXiv:2006.09311}},
  \href {https://doi.org/10.1088/1748-0221/15/12/P12038}
  {\path{doi:10.1088/1748-0221/15/12/P12038}}.

\bibitem{akerib2016low}
D.~Akerib, S.~Alsum, H.~Ara{\'u}jo, X.~Bai, A.~Bailey, J.~Balajthy,
  P.~Beltrame, E.~Bernard, A.~Bernstein, T.~Biesiadzinski, et~al., Low-energy
  (0.7-74 kev) nuclear recoil calibration of the lux dark matter experiment
  using dd neutron scattering kinematics, arXiv preprint arXiv:1608.05381
  (2016).

\bibitem{aprile2019xenon1t}
E.~Aprile, J.~Aalbers, F.~Agostini, M.~Alfonsi, L.~Althueser, F.~Amaro, V.~C.
  Antochi, F.~Arneodo, L.~Baudis, B.~Bauermeister, et~al., Xenon1t dark matter
  data analysis: Signal and background models and statistical inference,
  Physical Review D 99~(11) (2019) 112009.

\bibitem{Luo:2023ebw}
L.~Luo, et~al., {Improvement on the Linearity Response of PandaX-4T with new
  Photomultiplier Tubes Bases} (12 2023).
\newblock \href {http://arxiv.org/abs/2401.00373} {\path{arXiv:2401.00373}}.

\bibitem{Sakamoto:2023ond}
S.~Sakamoto, T.~Hasegawa, Y.~Itow, S.~Kazama, M.~Kobayashi, M.~Yamashita,
  {Development of a low-noise SiPM for the DARWIN experiment}, PoS ICRC2023
  (2023) 1435.
\newblock \href {https://doi.org/10.22323/1.444.1435}
  {\path{doi:10.22323/1.444.1435}}.

\bibitem{Barrow:2016doe}
P.~Barrow, et~al., {Qualification Tests of the R11410-21 Photomultiplier Tubes
  for the XENON1T Detector}, JINST 12~(01) (2017) P01024.
\newblock \href {http://arxiv.org/abs/1609.01654} {\path{arXiv:1609.01654}},
  \href {https://doi.org/10.1088/1748-0221/12/01/P01024}
  {\path{doi:10.1088/1748-0221/12/01/P01024}}.

\bibitem{PandaX_FADC_2021}
C.~He, et~al., \href{https://doi.org/10.1088/1748-0221/16/12/t12015}{A 500
  {MS}/s waveform digitizer for {PandaX} dark matter experiments}, Journal of
  Instrumentation 16~(12) (2021) T12015.
\newblock \href {https://doi.org/10.1088/1748-0221/16/12/t12015}
  {\path{doi:10.1088/1748-0221/16/12/t12015}}.
\newline\urlprefix\url{https://doi.org/10.1088/1748-0221/16/12/t12015}

\bibitem{Zhou:2023vmz}
Y.~Zhou, X.~Chen, {Data reduction strategy in the PandaX-4T experiment} (11
  2023).
\newblock \href {http://arxiv.org/abs/2311.12412} {\path{arXiv:2311.12412}}.

\bibitem{Wang:2023wrr}
X.~Wang, Z.~Lei, Y.~Ju, J.~Liu, N.~Zhou, Y.~Chen, Z.~Wang, X.~Cui, Y.~Meng,
  L.~Zhao, {Design, construction and commissioning of the PandaX-30T liquid
  xenon management system}, JINST 18~(05) (2023) P05028.
\newblock \href {http://arxiv.org/abs/2301.06044} {\path{arXiv:2301.06044}},
  \href {https://doi.org/10.1088/1748-0221/18/05/P05028}
  {\path{doi:10.1088/1748-0221/18/05/P05028}}.

\bibitem{Distillation:2017}
E.~Aprile, et~al.,
  \href{https://link.springer.com/article/10.1140/epjc/s10052-017-5326-3}{The
  xenon1t dark matter experiment}, The European Physical Journal C 77 (2017)
  881.
\newblock \href {https://doi.org/10.1140/epjc/s10052-017-5326-3}
  {\path{doi:10.1140/epjc/s10052-017-5326-3}}.
\newline\urlprefix\url{https://link.springer.com/article/10.1140/epjc/s10052-017-5326-3}

\bibitem{Barber}
B.~N. Inc.
\newblock \href{https://barber-nichols.com}{[link]}.
\newline\urlprefix\url{https://barber-nichols.com}

\bibitem{Gong:2012thh}
H.~Gong, K.~L. Giboni, X.~Ji, A.~Tan, L.~Zhao, {The Cryogenic System for the
  Panda-X Dark Matter Search Experiment}, JINST 8 (2013) P01002.
\newblock \href {http://arxiv.org/abs/1207.5100} {\path{arXiv:1207.5100}},
  \href {https://doi.org/10.1088/1748-0221/8/01/P01002}
  {\path{doi:10.1088/1748-0221/8/01/P01002}}.

\bibitem{Zhao:2020vxh}
L.~Zhao, X.~Cui, W.~Ma, Y.~Fan, K.~Giboni, T.~Zhang, J.~Liu, X.~Ji, {The
  cryogenics and xenon handling system for the PandaX-4T experiment}, JINST
  16~(06) (2021) T06007.
\newblock \href {http://arxiv.org/abs/2012.10583} {\path{arXiv:2012.10583}},
  \href {https://doi.org/10.1088/1748-0221/16/06/T06007}
  {\path{doi:10.1088/1748-0221/16/06/T06007}}.

\bibitem{ThermalManagement}
T.~Zhang, J.~Liu, et~al., {Thermal Management Design and Key Technology
  Validation for PandaX Underground Experiment}\href
  {https://doi.org/10.48550/arXiv.2408.13433}
  {\path{doi:10.48550/arXiv.2408.13433}}.

\bibitem{Giboni:2019fqo}
K.~L. Giboni, P.~Juyal, E.~Aprile, Y.~Zhang, J.~Naganoma, {A LN$_{2}$-based
  cooling system for a next-generation liquid xenon dark matter detector},
  Nucl. Sci. Tech. 31~(8) (2020) 76.
\newblock \href {http://arxiv.org/abs/1909.09698} {\path{arXiv:1909.09698}},
  \href {https://doi.org/10.1007/s41365-020-00786-7}
  {\path{doi:10.1007/s41365-020-00786-7}}.

\bibitem{Piston1}
F.~LePort, et~al., {A Magnetically-driven piston pump for ultra-clean
  applications}, Rev. Sci. Instrum. 82 (2011) 105114.
\newblock \href {http://arxiv.org/abs/1104.5041} {\path{arXiv:1104.5041}},
  \href {https://doi.org/10.1063/1.3653391} {\path{doi:10.1063/1.3653391}}.

\bibitem{Piston2}
E.~Brown, et~al., {Magnetically-coupled piston pump for high-purity gas
  applications}, Eur. Phys. J. C 78~(7) (2018) 604.
\newblock \href {http://arxiv.org/abs/1803.08498} {\path{arXiv:1803.08498}},
  \href {https://doi.org/10.1140/epjc/s10052-018-6062-z}
  {\path{doi:10.1140/epjc/s10052-018-6062-z}}.

\bibitem{XENON:2022ltv}
E.~Aprile, et~al., {Search for New Physics in Electronic Recoil Data from
  XENONnT}, Phys. Rev. Lett. 129~(16) (2022) 161805.
\newblock \href {http://arxiv.org/abs/2207.11330} {\path{arXiv:2207.11330}},
  \href {https://doi.org/10.1103/PhysRevLett.129.161805}
  {\path{doi:10.1103/PhysRevLett.129.161805}}.

\bibitem{Liquid_purity:2022}
G.~Plante, E.~Aprile, J.~Howlett, Y.~Zhang, {Liquid-phase purification for
  multi-tonne xenon detectors}, Eur. Phys. J. C 82~(10) (2022) 860.
\newblock \href {http://arxiv.org/abs/2205.07336} {\path{arXiv:2205.07336}},
  \href {https://doi.org/10.1140/epjc/s10052-022-10832-w}
  {\path{doi:10.1140/epjc/s10052-022-10832-w}}.

\bibitem{Distillation:JINST2014}
Z.~Wang, et~al.,
  \href{https://iopscience.iop.org/article/10.1088/1748-0221/9/11/P11024}{Large
  scale xenon purification using cryogenic distillation for dark matter
  detectors}, Journal of Instrumentation 9 (2014) P11024.
\newblock \href {https://doi.org/10.1088/1748-0221/9/11/P11024}
  {\path{doi:10.1088/1748-0221/9/11/P11024}}.
\newline\urlprefix\url{https://iopscience.iop.org/article/10.1088/1748-0221/9/11/P11024}

\bibitem{Distillation:JINST2021}
X.~Cui, et~al.,
  \href{https://iopscience.iop.org/article/10.1088/1748-0221/16/07/P07046}{Design
  and commissioning of the pandax-4t cryogenic distillation system for krypton
  and radon removal}, Journal of Instrumentation 16 (2021) P07046.
\newblock \href {https://doi.org/10.1088/1748-0221/16/07/P07046}
  {\path{doi:10.1088/1748-0221/16/07/P07046}}.
\newline\urlprefix\url{https://iopscience.iop.org/article/10.1088/1748-0221/16/07/P07046}

\bibitem{Distillation:JINST2021-2}
R.~Yan, et~al.,
  \href{https://pubs.aip.org/aip/rsi/article/92/12/123303/283310/PandaX-4T-cryogenic-distillation-system-for}{Pandax-4t
  cryogenic distillation system for removing krypton from xenon}, Review of
  Scientific Instruments 92 (2021) 123303.
\newblock \href {https://doi.org/10.1063/5.0065154}
  {\path{doi:10.1063/5.0065154}}.
\newline\urlprefix\url{https://pubs.aip.org/aip/rsi/article/92/12/123303/283310/PandaX-4T-cryogenic-distillation-system-for}

\bibitem{Distillation:JINST2023}
X.~Cui, et~al.,
  \href{https://iopscience.iop.org/article/10.1088/1748-0221/19/07/P07010}{Radon
  removal commissioning of the pandax-4t cryogenic distillation system},
  Journal of Instrumnetation 19 (2024) P07010.
\newblock \href {https://doi.org/10.1088/1748-0221/19/07/P07010}
  {\path{doi:10.1088/1748-0221/19/07/P07010}}.
\newline\urlprefix\url{https://iopscience.iop.org/article/10.1088/1748-0221/19/07/P07010}

\bibitem{Distillation:2012}
E.~Aprile, et~al.,
  \href{https://www.sciencedirect.com/science/article/pii/S0927650512000059}{The
  xenon100 dark matter experiment}, Astroparticle Physics, 2012, 35(9), 573-590
  35~(9) (2012) 573--590.
\newblock \href {https://doi.org/10.1016/j.astropartphys.2012.01.003}
  {\path{doi:10.1016/j.astropartphys.2012.01.003}}.
\newline\urlprefix\url{https://www.sciencedirect.com/science/article/pii/S0927650512000059}

\bibitem{MAJORANA:2016lsk}
N.~Abgrall, et~al., {The Majorana Demonstrator radioassay program}, Nucl.
  Instrum. Meth. A 828 (2016) 22--36.
\newblock \href {http://arxiv.org/abs/1601.03779} {\path{arXiv:1601.03779}},
  \href {https://doi.org/10.1016/j.nima.2016.04.070}
  {\path{doi:10.1016/j.nima.2016.04.070}}.

\bibitem{PhysRevC.92.015503}
J.~B. Albert, et~al.,
  \href{https://link.aps.org/doi/10.1103/PhysRevC.92.015503}{Investigation of
  radioactivity-induced backgrounds in exo-200}, Phys. Rev. C 92 (2015) 015503.
\newblock \href {https://doi.org/10.1103/PhysRevC.92.015503}
  {\path{doi:10.1103/PhysRevC.92.015503}}.
\newline\urlprefix\url{https://link.aps.org/doi/10.1103/PhysRevC.92.015503}

\bibitem{PhysRevD.95.082002}
R.~Agnese, et~al.,
  \href{https://link.aps.org/doi/10.1103/PhysRevD.95.082002}{Projected
  sensitivity of the supercdms snolab experiment}, Phys. Rev. D 95 (2017)
  082002.
\newblock \href {https://doi.org/10.1103/PhysRevD.95.082002}
  {\path{doi:10.1103/PhysRevD.95.082002}}.
\newline\urlprefix\url{https://link.aps.org/doi/10.1103/PhysRevD.95.082002}

\bibitem{XENON:2022evz}
E.~Aprile, et~al., {Double-Weak Decays of $^{124}$Xe and $^{136}$Xe in the
  XENON1T and XENONnT Experiments}, Phys. Rev. C 106~(2) (2022) 024328.
\newblock \href {http://arxiv.org/abs/2205.04158} {\path{arXiv:2205.04158}},
  \href {https://doi.org/10.1103/PhysRevC.106.024328}
  {\path{doi:10.1103/PhysRevC.106.024328}}.

\bibitem{DARWIN:2020jme}
F.~Agostini, et~al., {Sensitivity of the DARWIN observatory to the neutrinoless
  double beta decay of $^{136}$Xe}, Eur. Phys. J. C 80~(9) (2020) 808,
  [Erratum: Eur. Phys. J. C. 83, 996 (2023)].
\newblock \href {http://arxiv.org/abs/2003.13407} {\path{arXiv:2003.13407}},
  \href {https://doi.org/10.1140/epjc/s10052-020-8196-z}
  {\path{doi:10.1140/epjc/s10052-020-8196-z}}.

\bibitem{JNE:2020bwn}
Z.~Guo, et~al., {Muon flux measurement at China Jinping Underground
  Laboratory}, Chin. Phys. C 45~(2) (2021) 025001.
\newblock \href {http://arxiv.org/abs/2007.15925} {\path{arXiv:2007.15925}},
  \href {https://doi.org/10.1088/1674-1137/abccae}
  {\path{doi:10.1088/1674-1137/abccae}}.

\bibitem{Chen:2021asx}
X.~Chen, et~al.,
  \href{http://dx.doi.org/10.1088/1748-0221/16/09/T09004}{{BambooMC}
  {\textemdash} a {Geant4} simulation program for the {PandaX} experiments},
  Journal of Instrumentation 16~(09) (2021) T09004.
\newblock \href {https://doi.org/10.1088/1748-0221/16/09/t09004}
  {\path{doi:10.1088/1748-0221/16/09/t09004}}.
\newline\urlprefix\url{http://dx.doi.org/10.1088/1748-0221/16/09/T09004}

\bibitem{osti_15215}
D.~G. Madland, E.~D. Arthur, G.~P. Estes, J.~E. Stewart, M.~Bozoian, R.~T.
  Perry, T.~A. Parish, T.~H. Brown, T.~R. England, W.~B. Wilson, W.~S.
  Charlton, \href{https://www.osti.gov/biblio/15215}{Sources 4a: A code for
  calculating (alpha,n), spontaneous fission, and delayed neutron sources and
  spectra} (9 1999).
\newblock \href {https://doi.org/10.2172/15215} {\path{doi:10.2172/15215}}.
\newline\urlprefix\url{https://www.osti.gov/biblio/15215}

\bibitem{PandaX-II:2019jmf}
Q.~Wang, et~al., {An Improved Evaluation of the Neutron Background in the
  PandaX-II Experiment}, Sci. China Phys. Mech. Astron. 63~(3) (2020) 231011.
\newblock \href {http://arxiv.org/abs/1907.00545} {\path{arXiv:1907.00545}},
  \href {https://doi.org/10.1007/s11433-019-9603-9}
  {\path{doi:10.1007/s11433-019-9603-9}}.

\bibitem{PandaX:2024jjs}
X.~Lu, et~al., {Measurement of solar pp neutrino flux using electron recoil
  data from PandaX-4T commissioning run*}, Chin. Phys. C 48~(9) (2024) 091001.
\newblock \href {http://arxiv.org/abs/2401.07045} {\path{arXiv:2401.07045}},
  \href {https://doi.org/10.1088/1674-1137/ad582a}
  {\path{doi:10.1088/1674-1137/ad582a}}.

\bibitem{MSzydagis_2011}
M.~Szydagis, N.~Barry, K.~Kazkaz, J.~Mock, D.~Stolp, M.~Sweany, M.~Tripathi,
  S.~Uvarov, N.~Walsh, M.~Woods,
  \href{https://dx.doi.org/10.1088/1748-0221/6/10/P10002}{Nest: a comprehensive
  model for scintillation yield in liquid xenon}, Journal of Instrumentation
  6~(10) (2011) P10002.
\newblock \href {https://doi.org/10.1088/1748-0221/6/10/P10002}
  {\path{doi:10.1088/1748-0221/6/10/P10002}}.
\newline\urlprefix\url{https://dx.doi.org/10.1088/1748-0221/6/10/P10002}

\bibitem{szydagis2018noble}
M.~Szydagis, J.~Balajthy, J.~Brodsky, J.~Cutter, J.~Huang, E.~Kozlova,
  B.~Lenardo, A.~Manalaysay, D.~McKinsey, M.~Mooney, et~al., Noble element
  simulation technique v2. 0, Zenodo: Geneve, Switzerland (2018).

\bibitem{Vinyoles:2016djt}
N.~Vinyoles, A.~M. Serenelli, F.~L. Villante, S.~Basu, J.~Bergstr\"om, M.~C.
  Gonzalez-Garcia, M.~Maltoni, C.~Pe\~na Garay, N.~Song, {A new Generation of
  Standard Solar Models}, Astrophys. J. 835~(2) (2017) 202.
\newblock \href {http://arxiv.org/abs/1611.09867} {\path{arXiv:1611.09867}},
  \href {https://doi.org/10.3847/1538-4357/835/2/202}
  {\path{doi:10.3847/1538-4357/835/2/202}}.

\bibitem{aalbers2022next}
J.~Aalbers, S.~AbdusSalam, K.~Abe, V.~Aerne, F.~Agostini, S.~A. Maouloud,
  D.~Akerib, D.~Y. Akimov, J.~Akshat, A.~Al~Musalhi, et~al., A next-generation
  liquid xenon observatory for dark matter and neutrino physics, Journal of
  Physics G: Nuclear and Particle Physics 50~(1) (2022) 013001.

\bibitem{Macolino:2020uqq}
C.~Macolino, {DARWIN: direct dark matter search with the ultimate detector}, J.
  Phys. Conf. Ser. 1468~(1) (2020) 012068.
\newblock \href {https://doi.org/10.1088/1742-6596/1468/1/012068}
  {\path{doi:10.1088/1742-6596/1468/1/012068}}.

\bibitem{Baxter:2021pqo}
D.~Baxter, et~al., {Recommended conventions for reporting results from direct
  dark matter searches}, Eur. Phys. J. C 81~(10) (2021) 907.
\newblock \href {http://arxiv.org/abs/2105.00599} {\path{arXiv:2105.00599}},
  \href {https://doi.org/10.1140/epjc/s10052-021-09655-y}
  {\path{doi:10.1140/epjc/s10052-021-09655-y}}.

\bibitem{Bagnaschi:2018}
E.~Bagnaschi, K.~Sakurai, M.~Borsato, O.~Buchmueller, M.~Citron, J.~C. Costa,
  A.~De~Roeck, M.~J. Dolan, J.~R. Ellis, H.~Flächer, S.~Heinemeyer, M.~Lucio,
  D.~Martínez~Santos, K.~A. Olive, A.~Richards, V.~C. Spanos,
  I.~Suárez~Fernández, G.~Weiglein, {Likelihood analysis of the pMSSM11 in
  light of LHC 13-TeV data}, The European Physical Journal C 78~(3) (Mar.
  2018).
\newblock \href {https://doi.org/10.1140/epjc/s10052-018-5697-0}
  {\path{doi:10.1140/epjc/s10052-018-5697-0}}.

\bibitem{PandaX:2022aac}
W.~Ma, et~al., {Search for Solar B8 Neutrinos in the PandaX-4T Experiment Using
  Neutrino-Nucleus Coherent Scattering}, Phys. Rev. Lett. 130~(2) (2023)
  021802.
\newblock \href {http://arxiv.org/abs/2207.04883} {\path{arXiv:2207.04883}},
  \href {https://doi.org/10.1103/PhysRevLett.130.021802}
  {\path{doi:10.1103/PhysRevLett.130.021802}}.

\bibitem{XENON1T2018SI}
E.~Aprile, et~al., {Dark Matter Search Results from a One Ton-Year Exposure of
  XENON1T}, Phys. Rev. Lett. 121~(11) (2018) 111302.
\newblock \href {http://arxiv.org/abs/1805.12562} {\path{arXiv:1805.12562}},
  \href {https://doi.org/10.1103/PhysRevLett.121.111302}
  {\path{doi:10.1103/PhysRevLett.121.111302}}.

\bibitem{LUX2017SI}
D.~S. Akerib, et~al., {Results from a search for dark matter in the complete
  LUX exposure}, Phys. Rev. Lett. 118~(2) (2017) 021303.
\newblock \href {http://arxiv.org/abs/1608.07648} {\path{arXiv:1608.07648}},
  \href {https://doi.org/10.1103/PhysRevLett.118.021303}
  {\path{doi:10.1103/PhysRevLett.118.021303}}.

\bibitem{xenon1t_sd_2019}
E.~Aprile, et~al., {Constraining the spin-dependent WIMP-nucleon cross sections
  with XENON1T}, Phys. Rev. Lett. 122~(14) (2019) 141301.
\newblock \href {http://arxiv.org/abs/1902.03234} {\path{arXiv:1902.03234}},
  \href {https://doi.org/10.1103/PhysRevLett.122.141301}
  {\path{doi:10.1103/PhysRevLett.122.141301}}.

\bibitem{lux_sd_2017}
D.~S. Akerib, et~al., {Limits on spin-dependent WIMP-nucleon cross section
  obtained from the complete LUX exposure}, Phys. Rev. Lett. 118~(25) (2017)
  251302.
\newblock \href {http://arxiv.org/abs/1705.03380} {\path{arXiv:1705.03380}},
  \href {https://doi.org/10.1103/PhysRevLett.118.251302}
  {\path{doi:10.1103/PhysRevLett.118.251302}}.

\bibitem{Baudis:2024_XLZD}
L.~Baudis, {DARWIN/XLZD: A future xenon observatory for dark matter and other
  rare interactions}, Nuclear Physics B 1003 (2024) 116473, special Issue of
  Nobel Symposium 182 on Dark Matter.
\newblock \href
  {https://doi.org/https://doi.org/10.1016/j.nuclphysb.2024.116473}
  {\path{doi:https://doi.org/10.1016/j.nuclphysb.2024.116473}}.

\bibitem{McDonald:2024osu}
A.~B. McDonald, {Dark matter detection with liquid argon}, Nucl. Phys. B 1003
  (2024) 116436.
\newblock \href {https://doi.org/10.1016/j.nuclphysb.2024.116436}
  {\path{doi:10.1016/j.nuclphysb.2024.116436}}.

\bibitem{Dolinski:2019nrj}
M.~J. Dolinski, A.~W.~P. Poon, W.~Rodejohann, {Neutrinoless Double-Beta Decay:
  Status and Prospects}, Ann. Rev. Nucl. Part. Sci. 69 (2019) 219--251.
\newblock \href {http://arxiv.org/abs/1902.04097} {\path{arXiv:1902.04097}},
  \href {https://doi.org/10.1146/annurev-nucl-101918-023407}
  {\path{doi:10.1146/annurev-nucl-101918-023407}}.

\bibitem{Kumaran:2021lvv}
S.~Kumaran, L.~Ludhova, O.~Penek, G.~Settanta, {Borexino Results on Neutrinos
  from the Sun and Earth}, Universe 7~(7) (2021) 231.
\newblock \href {http://arxiv.org/abs/2105.13858} {\path{arXiv:2105.13858}},
  \href {https://doi.org/10.3390/universe7070231}
  {\path{doi:10.3390/universe7070231}}.

\bibitem{aharmim2013combined}
B.~Aharmim, S.~Ahmed, A.~Anthony, N.~Barros, E.~Beier, A.~Bellerive,
  B.~Beltran, M.~Bergevin, S.~Biller, K.~Boudjemline, et~al., Combined analysis
  of all three phases of solar neutrino data from the sudbury neutrino
  observatory, Physical Review C 88~(2) (2013) 025501.

\bibitem{vinyoles2017new}
N.~Vinyoles, A.~M. Serenelli, F.~L. Villante, S.~Basu, J.~Bergstr{\"o}m,
  M.~Gonzalez-Garcia, M.~Maltoni, C.~Pe{\~n}a-Garay, N.~Song, A new generation
  of standard solar models, The Astrophysical Journal 835~(2) (2017) 202.

\bibitem{Borexinodmumag}
M.~Agostini, et~al., {Limiting neutrino magnetic moments with Borexino Phase-II
  solar neutrino data}, Phys. Rev. D 96~(9) (2017) 091103.
\newblock \href {https://doi.org/10.1103/PhysRevD.96.091103}
  {\path{doi:10.1103/PhysRevD.96.091103}}.

\bibitem{Gemmamumag}
A.~Beda, V.~Brudanin, V.~Egorov, D.~Medvedev, V.~Pogosov, E.~Shevchik,
  M.~Shirchenko, A.~Starostin, I.~Zhitnikov, {Gemma experiment: The results of
  neutrino magnetic moment search}, Phys. Part. Nucl. Lett. 10 (2013) 139--143.
\newblock \href {https://doi.org/10.1134/S1547477113020027}
  {\path{doi:10.1134/S1547477113020027}}.

\bibitem{LZ:2023poo}
J.~Aalbers, et~al., {Search for new physics in low-energy electron recoils from
  the first LZ exposure}, Phys. Rev. D 108~(7) (2023) 072006.
\newblock \href {http://arxiv.org/abs/2307.15753} {\path{arXiv:2307.15753}},
  \href {https://doi.org/10.1103/PhysRevD.108.072006}
  {\path{doi:10.1103/PhysRevD.108.072006}}.

\bibitem{WhiteDwarfmumag}
M.~M. Miller~Bertolami, {Limits on the neutrino magnetic dipole moment from the
  luminosity function of hot white dwarfs}, Astron. Astrophys. 562 (2014) A123.
\newblock \href {https://doi.org/10.1051/0004-6361/201322641}
  {\path{doi:10.1051/0004-6361/201322641}}.

\bibitem{GlobularClustermumag}
S.~A. Díaz, K.-P. Schröder, K.~Zuber, D.~Jack, E.~E.~B. Barrios, {Constraint
  on the axion-electron coupling constant and the neutrino magnetic dipole
  moment by using the tip-RGB luminosity of fifty globular clusters} (10 2019).
\newblock \href {http://arxiv.org/abs/1910.10568} {\path{arXiv:1910.10568}}.

\bibitem{PhysRevB.27.5279}
A.~Hitachi, T.~Takahashi, N.~Funayama, K.~Masuda, J.~Kikuchi, T.~Doke,
  \href{https://link.aps.org/doi/10.1103/PhysRevB.27.5279}{Effect of ionization
  density on the time dependence of luminescence from liquid argon and xenon},
  Phys. Rev. B 27 (1983) 5279--5285.
\newblock \href {https://doi.org/10.1103/PhysRevB.27.5279}
  {\path{doi:10.1103/PhysRevB.27.5279}}.
\newline\urlprefix\url{https://link.aps.org/doi/10.1103/PhysRevB.27.5279}

\bibitem{LUX:2018zdm}
D.~S. Akerib, et~al., {Liquid xenon scintillation measurements and pulse shape
  discrimination in the LUX dark matter detector}, Phys. Rev. D 97~(11) (2018)
  112002.
\newblock \href {http://arxiv.org/abs/1802.06162} {\path{arXiv:1802.06162}},
  \href {https://doi.org/10.1103/PhysRevD.97.112002}
  {\path{doi:10.1103/PhysRevD.97.112002}}.

\bibitem{Hogenbirk:2018zwf}
E.~Hogenbirk, J.~Aalbers, P.~A. Breur, M.~P. Decowski, K.~van Teutem, A.~P.
  Colijn, {Precision measurements of the scintillation pulse shape for
  low-energy recoils in liquid xenon}, JINST 13~(05) (2018) P05016.
\newblock \href {http://arxiv.org/abs/1803.07935} {\path{arXiv:1803.07935}},
  \href {https://doi.org/10.1088/1748-0221/13/05/P05016}
  {\path{doi:10.1088/1748-0221/13/05/P05016}}.

\bibitem{Brodsky:2018abk}
J.~P. Brodsky, S.~Sangiorgio, M.~Heffner, T.~Stiegler, {Background
  Discrimination for Neutrinoless Double Beta Decay in Liquid Xenon Using
  Cherenkov Light}, Nucl. Instrum. Meth. A 922 (2019) 76--83.
\newblock \href {http://arxiv.org/abs/1812.05694} {\path{arXiv:1812.05694}},
  \href {https://doi.org/10.1016/j.nima.2018.12.057}
  {\path{doi:10.1016/j.nima.2018.12.057}}.

\bibitem{Honda:2015fha}
M.~Honda, M.~Sajjad~Athar, T.~Kajita, K.~Kasahara, S.~Midorikawa, {Atmospheric
  neutrino flux calculation using the NRLMSISE-00 atmospheric model}, Phys.
  Rev. D 92~(2) (2015) 023004.
\newblock \href {http://arxiv.org/abs/1502.03916} {\path{arXiv:1502.03916}},
  \href {https://doi.org/10.1103/PhysRevD.92.023004}
  {\path{doi:10.1103/PhysRevD.92.023004}}.

\bibitem{Suzuki:2000ch}
Y.~Suzuki, {Low-energy solar neutrino detection by using liquid xenon}, in:
  {Workshop on Solar Neutrinos below 1-MeV: NuLow}, 2000.
\newblock \href {http://arxiv.org/abs/hep-ph/0008296}
  {\path{arXiv:hep-ph/0008296}}, \href
  {https://doi.org/10.1142/9789812778000_0009}
  {\path{doi:10.1142/9789812778000_0009}}.

\bibitem{Schumann:2015cpa}
M.~Schumann, L.~Baudis, L.~B\"utikofer, A.~Kish, M.~Selvi, {Dark matter
  sensitivity of multi-ton liquid xenon detectors}, JCAP 10 (2015) 016.
\newblock \href {http://arxiv.org/abs/1506.08309} {\path{arXiv:1506.08309}},
  \href {https://doi.org/10.1088/1475-7516/2015/10/016}
  {\path{doi:10.1088/1475-7516/2015/10/016}}.

\end{thebibliography}

\end{document}